\documentclass[12pt]{article} \textheight= 240mm \textwidth=160mm \topmargin = -1 cm
\usepackage{graphicx}
\usepackage{amssymb}

\newcommand{\mm}{m}

\input epsf

\begin{document}

\title{\textbf{Geometry of irreversibility: The film of nonequilibrium states}}

\author{Alexander N. Gorban$^{1,2,3}$\thanks{agorban$@$mat.ethz.ch, $^{**}$ikarlin$@$mat.ethz.ch},
\and Iliya V. Karlin$^{1,2**}\!\!$, \\ $^{1}$ ETH-Zentrum, Department of Materials, Institute of
Polymers, \\ Sonneggstr. 3, ML J19, CH-8092 Z{\"u}rich, Switzerland; \\ $^{2}$ Institute of
Computational Modeling SB RAS, \\ Akademgorodok, Krasnoyarsk 660036, Russia; \\ $^{3}$ Institut des
Hautes Etudes Scientifiques, \\ Le Bois-Marie, 35, route de Chartres, F-91440, Bures-sur-Yvette,
France}

\date{}

\maketitle


\begin{abstract}

A general geometrical framework of nonequilibrium thermodynamics is developed. The notion of {\it
macroscopically definable ensembles} is developed. The thesis about macroscopically definable
ensembles is suggested. This thesis should play  the same role in the nonequilibrium thermodynamics,
as the Church-Turing thesis in the theory of computability. The {\it  primitive macroscopically
definable ensembles} are described. These are ensembles with macroscopically prepared initial states.
The method for computing trajectories of primitive macroscopically definable nonequilibrium ensembles
is elaborated. These trajectories are represented as sequences of deformed equilibrium ensembles and
simple quadratic models between them. The primitive macroscopically definable ensembles form the
manifold in the space of ensembles. We call this manifold the {\it film of nonequilibrium states}.
The equation for the film and the equation for the ensemble motion on the film are written down. The
notion of the invariant film of non-equilibrium states, and the method of its approximate
construction transform the the problem of nonequilibrium kinetics into a series of problems of
equilibrium statistical physics. The developed methods allow us to solve the problem of
macro-kinetics even when there are no autonomous equations of macro-kinetics.

\end{abstract}

\clearpage

\tableofcontents

\clearpage

\addcontentsline{toc}{section}{\textbf{Introduction}}

\section*{Introduction}

The goal of this paper is to discuss the nonlinear problems of irreversibility, and to revise the
previous attempts to solve them. The interest to the problem of irreversibility persists during
decades. It has been intensively discussed in the past, and nice accounts of these discussions can be
found in the literature (see, for example, \cite{Mont,GarCol,5,22}). Here, we intend to develop a
more geometrical viewpoint on the subject. The paper consists of two parts. First, in section
\ref{one}, we discuss in an informal way the origin of the problem, and demonstrate how the basic
constructions arise. Second, in section \ref{two}, we give a consistent geometric formalization of
these constructions. Our presentation is based on the notion of the natural projection introduced in
section \ref{natproj}. We discuss in detail the method of natural projector as the consistent
formalization of Ehrenfest's ideas of coarse-graining.

In section \ref{plenka} we introduce a one-dimensional model of nonequilibrium states. In the
background of many derivations of nonequilibrium kinetic equations there is present the following
picture: Above each point of the quasiequilibrium manifold there is located a huge subspace of
nonequilibrium distributions with the same values of the macroscopic variables, as in the
quasiequilibrium point. It is as if the motion decomposes into two projections, above the point on
the quasiequilibrium manifold, and in the projection on this manifold. The motion in each layer above
the points  is highly complicated, but fast, and everything quickly settles in this fast motion.

However, upon a more careful looking into the motions which start from the quasiequilibrium points,
we will observe that, above each point of the quasiequilibrium manifold it is located {\it just a
single curve,} and all the nonequilibrium (not-quasiequilibrium) states which come into the game form
just a one-dimensional manifold.

The novel approach developed in section \ref{natproj} allows to go beyond limitations  of the short
memory approximations through a study of stability of the quasiequilibrium manifold.

The one-dimensional models of nonequilibrium states form a {\it film of nonequilibrium states} over
the quasiequilibrium manifold. In section \ref{4} we present a collection of methods for the film
construction. One of the benefits from this new technic is the possibility to solve the problem of
macro-kinetic in cases when there are no autonomous equations of macro-kinetic for moment variables.
The notion of the invariant film of non-equilibrium states, and the method of its approximate
construction transform the the problem of nonequilibrium kinetics into a series of problems of
equilibrium statistical physics. To describe a  dynamics of nonequilibrium ensemble one should find
series of deformed equilibrium ensembles.

In Appendix a short presentation of the method of invariant manifold for kinetic problems is given.

The most important results of the paper are:

\begin{enumerate}
\item{The notion of {\it macroscopically definable ensembles} is developed.}
\item{ The {\it  primitive macroscopically definable ensembles} are described.}
\item{The method for computing trajectories of primitive macroscopically definable
nonequilibrium ensembles is elaborated. These trajectories are represented as series of deformed
equilibrium ensembles and simple quadratic models between them.}
\end{enumerate}

Let us give here an introductory description of these results.

The notion of macroscopically definable ensembles consists of three ingredients:

\begin{enumerate}
\item{Macroscopic variables, the variables which values {\it can be controlled by us}};
\item{Quasiequilibrium state, the conditional equilibrium state for fixed values of the
macroscopic variables;}
\item{Natural dynamics of the system.}
\end{enumerate}

We use the simplest representation of the control: For some moments of time we fix some values of the
macroscopic variables (of all these variables, or of part of them; for the whole system, or for
macroscopically defined part of it; the current ``natural", or some arbitrary value of these
variables), and the system obtains corresponding conditional equilibrium state. We can also keep
fixed values of some macroscopic variables during a time interval.

These control operations are discrete in time. The continuous control can be created by the closure:
the limit of a sequence of macroscopically definable ensembles is macroscopically definable too.

The role of the macroscopic variables for the irreversibility problem became clear to M. Leontovich
and J. Lebowitz several decades ago \cite{Leon,Leb,Leb1,Leb2,Leb3}. But this was the first step. Now
we do need the elaborate notion of ensembles which can be obtained by macroscopic tools. The Maxwell
Demon gives  the first clear picture of a difference between macroscopic and microscopic tools for
ensembles control (there are books devoted to analysis of this Demon  \cite{Max1,Max2}).
Nevertheless, the further step to the analysis of the notion of macroscopic definability in context
of constructive transition from microdynamics to macrokinetics equations had not done before the
paper \cite{20}. Our analysis pretends to be an analogue of the Church-Turing thesis
\cite{Chu1,Chu2}. This thesis concerns the notion of an effective (or mechanical) method in
mathematics. As a ``working hypothesis", Church proposed: A function of positive integers is
effectively calculable only if recursive.

We introduce a class of ``macroscopically definable ensembles" and formulate the thesis: An ensemble
can be macroscopically obtained only if macroscopically definable in according to the introduced
notion. This is a thesis about success of the formalization, as the Church-Turing thesis, and nobody
can prove or disprove it in rigorous sense, as well, as this famous thesis.

Another important new notion is the {\it ``macroscopically definable transformation"} of the
ensemble: If one get an ensemble, how can he transform it? First, it is possible just let them
evolve, second, it can be controlled by the macroscopic tools on the defined way (it is necessary
just to keep values of some macroscopic variables during some time).

The {\it  primitive macroscopically definable ensembles} are ensembles with quasiequilibrium initial
states and without further macroscopic control. These ensembles are prepared macroscopically, and
evolve due to natural dynamics. The significance of this class of ensembles is determined by the {\it
hypothesis about the primitive macroscopically definable ensembles}: Any macroscopically definable
ensemble can be approximated by primitive macroscopically definable ensembles with appropriate
accuracy. Now we have no other effective way to decribe the nonequilibrium state.

The primitive macroscopically definable ensembles form the manifold in the space of ensembles. We
call this manifold the ``film of nonequilibrium states". The equation for the film and the equation
for the ensemble motion on the film are written down.

The film of nonequilibrium states is the trajectory of the manifold of initial quasiequilibrium
states due to the natural (microscopic) dynamics. For every value of macroscopic variables this film
gives us a curve. The curvature of this curve defines kinetic coefficients and entropy production.

The main technical problem is the computation of this curve for arbitrary value of the macroscopic
variables. We represent it as a sequence of special points and second-order polynomial (Kepler)
models for trajectory between these points. The method elaborated for the computation is the further
development of the method for initial layer problem in the Boltzmann kinetics \cite{10,11}. For
dissipative Boltzmann microkinetics it was sufficient to use the first-order models (with or without
smoothing). For conservative microkinetics it is necessary to use the highest-order models.
Application of this method to the lattice kinetic equations gave the following possibilities:

\begin{itemize}
\item{To create the Lattice-Boltzmann kinetics with $H$-theorem \cite{LB1};}
\item{To transform the Lattice-Boltzmann method into the numerically stable computational tool for
fluid flows and other dissipative systems out of equilibrium \cite{LB2};}
\item{To develop the Entropic Lattice Boltzmann method as a starting basis for the
formulation of a new class of turbulence models based on genuinely kinetic principles \cite{LB3}.}
\end{itemize}

Now we extend the method elaborated for dissipative systems \cite{10,11} to the higher-order models
for conservative systems. The iteration method for improvement of obtained approximations is proposed
too. It is a version of the Method of invariant manifold for kinetic problems, developed in the
series of papers \cite{3,8} (the almost exhaustive review of these works can be find in the paper
\cite{CMIM}) . The summary of this method is given in Appendix.

The constructing of the method of physically consistent computation is the central part of our paper.
It is neither a philosophical opus, nor only discussion of foundations of science.

The main results of this paper were presented in the talk given on the First Mexican Meeting on
Mathematical and Experimental Physics, Mexico City, September 10-14, 2001, and  in the lectures given
on the V Russian National Seminar ``Modeling of Nonequilibrium systems", Krasnoyarsk, October 18-20,
2002 \cite{Lec}.

\section{The problem of irreversibility}
\label{one}

\subsection{The phenomenon of the macroscopic irreversibility}

\begin{figure}[p]
\begin{centering}
\includegraphics[width=150mm, height=225mm]{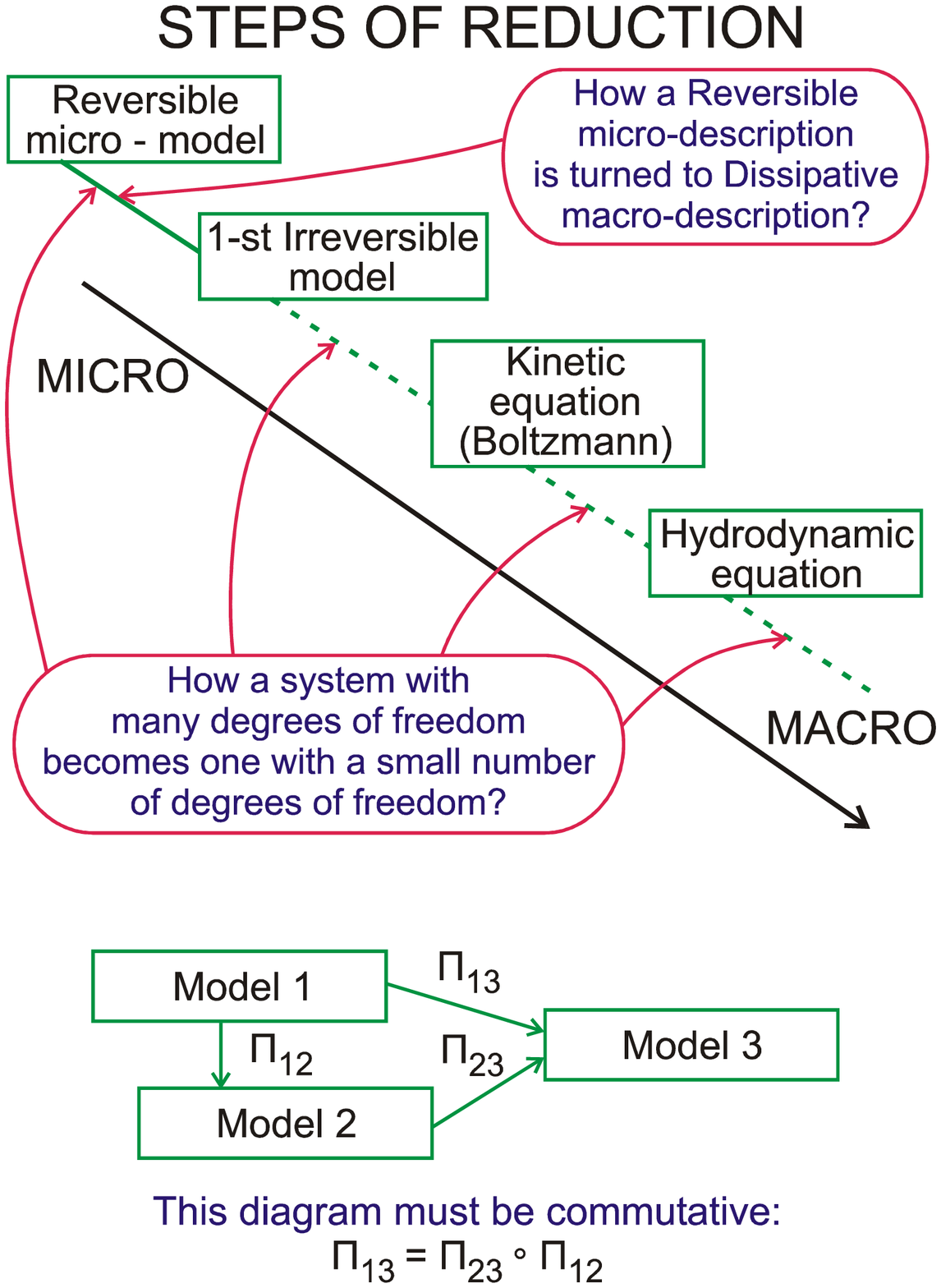}
\caption {The stairs of reduction, step by step.}
    \label{Figstaps}
\end{centering}
\end{figure}

The ``stairs of reduction" (Fig. \ref{Figstaps}) lead from the reversible microdynamics to
irreversible macrokinetics. The most mysterious is the first step: the appearance of irreversibility.

The best way to demonstrate the problem of irreversibility is the following {\it Gedankenexperiment}:
Let us watch the movie:  It's raining, people are running, cars rolling.  Let us now wind this movie
in the opposite direction, and we will see a strange and funny picture:  Drops of the rain are
raising up to the clouds, which next condensate into the vapor on the pools, on the surfaces of
rivers, people run with their backs forward, cars behave also quite strange, and so forth.  This
cannot be, and we ``know'' this for sure, we have never seen anything like this in our life.  Let us
now  imagine that we watch the same movie with a magnitude of $10^{8}-10^{9}$  so that we can resolve
individual particles.  And all of the sudden we discover that we cannot see any substantial
difference between the direct and the reverse demonstration:  Everywhere the particles are moving,
colliding, reacting according to the laws of physics, and nowhere there is a violation of anything.
We cannot tell the direct progressing of the  time from the reversed.  So, we have the
irreversibility of the macroscopic picture under the reversibility of the microscopic one.

Rain, people, cars - this all is  too complicated. One of the most  simple examples of the
irreversible macroscopic picture under the apparent reversibility of the microscopic picture (the
``thermal ratchet'')
 is given by R.\ Feynman
in his lectures on the character of physical law  \cite{Feinman}. We easily label  it as self-evident
the fact that particles of different colors mix together, and we would see it as a wonder the reverse
picture of a spontaneous decomposition of their mixture.  However, itself an appreciation of one
picture as usual, and of the other as unusual and wonderful - this is not yet the physics.  It is
desirable to measure somehow this transition from  order to disorder.

\subsection{Phase volume and dynamics of ensembles}

Let there be $n$ blue and $n$ white particles in a box, and let the box is separated in two halves,
the left and the right.  Location of all the particles in the box is described by the assembly of
$2n$ vectors of locations of individual particles.  The set of all the assemblies is a ``box'' in the
$6n$-dimensional space.  A point in this $6n$-dimensional box describes a configuration.  The motion
of this point is defined by equations of mechanics.

``Order'' is the configuration in which the blue particles are all in the right half, and all the
white particles are in the left half.  The set of all such configurations has a rather small volume.
It makes only $(1/2)^{2n}$ of the total volume of the $6n$-dimensional box. If $n=10$, this is of the
order of one per million of the total volume.  It is practically unthinkable to get into such a
configuration by chance.  It is also highly improbable that, by forming more or less voluntary the
initial conditions, we can observe that the system becomes ordered by itself.  From this standpoint,
the motion goes from the states of ``order'' to the state of ``disorder'', just because there are
many more states of ``disorder''.

However, we have defined it in this way.  The well known question of where is more order, in a fine
castle or in a pile of stones, has a profound answer:  It depends on which pile you mean.  If
``piles'' are thought as all configurations of stones which are not castles, then there are many more
such piles, and so there is less order in such a pile. However, if these are specially and uniquely
placed stones (for example, a garden of stones), then there is the same amount of order in such a
pile  as in the fine castle. {\it Not a  specific  configuration is important but an assembly of
configurations embraced by one notion.}

This transition from single configurations to their assemblies (ensembles) play the pivotal  role in
the understanding of irreversibility:  The irreversible transition from the ordered configuration
(blue particles are on the right, white particles are on the left) to the disordered one occurs
simply because there are many more of the disordered (in the sense of the volume).  Here, strictly
speaking, we have to add also a reference to the Liouville theorem:  The volume in the phase space
which is occupied by the ensemble does not change in time as the mechanical system evolves.  Because
of this fact, the volume $V$ is a good measure to compare the assemblies of configurations.  However,
more often the quantity $\ln V$ is used, this is called the entropy.

The point which represents the configuration, very rapidly leaves a small neighborhood and for a long
time (in practice, never) does not come back into it. In this, seemingly idyllic picture, there are
still two rather dark  clouds left.  First, the arrow of time has not appeared.  If we move from the
ordered initial state (separated particles) backwards in time, then everything will stay the same as
when we move forward in time, that is, the order will be changing into the disorder.  Second, let us
wind the film backwards, let us shoot the movie about mixing of colored particles, and then let us
watch in the reverse order their demixing.  Then the initial configurations for the reverse motion
will only seem to be disordered.  Their ``order'' is in the fact that they were obtained from the
separated mixture by letting the system to evolve for the time $t$.  There are also very few such
configurations, just the same number as of the ordered (separated particles) states.  If we start
with these configurations, then we obtain the ordered system after the time $t$.  Then why this most
obvious consequence of the laws of mechanics looks so improbable on the screen?  Perhaps, it should
be accepted that states which are obtained from the ordered state by a time shift, and by inversion
of particle's velocities (in order to initialize the reverse motion in time), cannot be prepared by
using macroscopic means of preparation.  In order to prepare such states, one would have to employ an
army of Maxwell's daemons which would invert individual velocities with sufficient accuracy (here, it
is much more into the phrase ``sufficient accuracy''  but this has to be discussed separately and
next time).

For this reason, we lump the distinguished initial conditions, for which the mixture decomposes
spontaneously (``piles'' of special form, or ``gardens of stones'') together with other
configurations into {\it macroscopically definable ensembles}.  And already for those ensembles the
spontaneous demixing becomes improbable.  This way we come to a new viewpoint:  (i). We cannot
prepare individual systems but only representatives of ensembles. (ii) We cannot prepare ensembles at
our will but only ``macroscopically definable ensembles''. What are these macroscopically definable
ensembles?  It seems that  one has to give some constructions, the universality of which can  only be
proven by the  time and experience.

\subsection{Macroscopically definable ensembles and quasiequilibria}

The main tool in the study of macroscopically definable ensembles is the notion of the macroscopic
variables, and of the quasiequilibria.  In the dynamics of the ensembles, the macroscopic variables
are defined as linear functionals (moments) of the density distribution of the ensemble.  Macroscopic
variables $M$ usually include hydrodynamic fields, density of particles, densities of momentum, and
density of the energy, also the list may include stress tensor, reaction rates and other quantities.
In the present context, it is solely important that the list the macroscopic variables is identified
for the system under consideration.

A single system is characterized by a single point $x$ in the phase space. The ensemble of the
systems is defined by the probability density $F$ on the phase space. Density $F$ must satisfy a set
of restrictions, the most important of which are: Nonnegativity, $F(x)\ge 0$, normalization,
\begin{equation}
\label{norm} \int_X F(x)dV(x)=1,
\end{equation}
and that the entropy is defined, that is, there exists the integral,
\begin{equation}
\label{S} S(F)=-\int_X F(x)\ln F(x) dV(x).
\end{equation}
(Function $F\ln F$ is continuously extended to zero values of $F$: $0\ln 0=0$). Here $dV(x)$ is the
invariant  measure (phase volume).

The quasiequilibrium ensemble describes the ``equilibrium under restrictions''.  It is assumed that
some external forcing keeps the given values of the macroscopic variables $M$, with this, ``all the
rest" comes the corresponding (generalized) canonic ensemble $F$ which is the solution to the
problem:
\begin{equation}
\label{smax} S(F)\to \max,\ M(F)=M.
\end{equation}
 where $S(F)$ is the entropy, $M(F)$ is the set of macroscopic variables.

{\bf The thesis about the macroscopically definable ensembles}.  Macroscopically definable ensembles
are obtained as the result of two operations:

(i). Bringing the system into the quasiequilibrium state corresponding to either the whole set of the
macroscopic variables $M$, or to its subset.

(ii). Changing the ensemble according to the microscopic dynamics (due to the Liouville equation)
during some time $t$.

These operations can be applied in the interchanging order any number of times, and for arbitrary
time segments $t$. The limit of macroscopically definable ensembles will also be termed the
macroscopically definable. One always starts with the operation (i).

In order to work out the notion of macroscopic definability, one has to pay more attention to
partitioning the system into subsystems. This involves a partition of the phase space $X$ with the
measure $dV$ on it into a direct product of spaces, $X=X_1\times X_2$ with the measure $dV_1dV_2$. To
each admissible (``macroscopic") partition  into sub-systems, it corresponds the operation of taking
a ``partial quasiequilibrium'', applied to some density $F_0(x_1,x_2)$:
\begin{eqnarray}
\label{partial} &&S(F)\to {\rm max},\\\nonumber &&M(F)=M,\
\int_{X_2}F(x_1,x_2)dV_2(x_2)=\int_{X_2}F_0(x_1,x_2)dV_2(x_2).
\end{eqnarray}
where $M$ is some subset of macroscopic variables (not necessarily the whole list of the macroscopic
variables). In Eq.\ (\ref{partial}), the state of the first subsystem is not changing, whereas the
second subsystem is brought into the quasiequilibrium. In fact, the problem (\ref{partial}) is a
version of the problem (\ref{smax}) with additional ``macroscopic variables'',
\begin{equation}
\label{additional} \int_{X_2}F(x_1,x_2)dV_2(x_2).
\end{equation}

The extended thesis about macroscopically definable ensembles allows to use  also operations
(\ref{partial}) with only one restriction: The initial state should be the ``true quasiequilibrium''
that is, macroscopic variables related to all possible partitions into subsystems should appear only
after the sequence of operations has started with the solution to the problem (\ref{smax}) for some
initial $M$. This does not exclude a possibility of including operators (\ref{additional}) into the
list of the basic macroscopic variables $M$. The standard example of such an inclusion are few-body
distribution functions treated as macroscopic variables in derivations of kinetic equations from the
Liouville equation.

Irreversibility is related to the choice of the initial conditions. The extended set of
macroscopically definable ensembles is thus given by three objects:

(i). The set of macroscopic variables $M$ which are linear (and, in an appropriate topology,
continuous) mappings of the space of distributions onto the space of values of the macroscopic
variables;

(ii). Macroscopically admissible partitions of the system into sub-systems;

(iii). Equations of microscopic dynamics (the Liouville equation, for example).

The choice of the macroscopic variables and of the macroscopically admissible partitions is a
distinguished topic. The main question is: what variables are under the macroscopic control? Here the
macroscopic variables are represented as formal elements of the construction, and the arbitrariness
is removed only at solving specific problems. Usual we can postulate some properties of macroscopic
variables, for example, symmetry with respect to any permutation of equal micro-particles.

We have discussed the {\it prepared} ensembles. But there is another statement of problem too: Let us
get an ensemble. The way how we get it may be different and unknown, for example, some demon or
oracle\footnote{In the theory of computation, if there is a device which could answer questions
beyond those that a Turing machine can answer, then it would be called oracle.} can give it to us.
How can we transform this ensemble by macroscopic tools? First, it is possible just let them evolve,
second, it can be controlled by the macroscopic tools on the defined way (it is necessary just to
keep values of some macroscopic variables during some time).

{\bf The thesis about the macroscopically definable transformation of ensembles}.  Macroscopically
definable transformation of ensembles are obtained as the result of two operations:

(i). Bringing the system into the quasiequilibrium state corresponding to either the whole set of the
macroscopic variables $M$, or to its subset.

(ii). Changing the ensemble according to the microscopic dynamics (due to the Liouville equation, for
example) during some time $t$.

These operations can be applied in the interchanging order any number of times, and for arbitrary
time segments $t$. The limit of macroscopically definable transformations will also be termed the
macroscopically definable. The main difference of this definition (macroscopically definable
transformation) from the definition of the macroscopically definable ensembles is the absence of
restriction on the initial state, one can start from arbitrary ensemble.

The class of macroscopically definable ensembles includes the more simple, but important class. Let
us reduce the macroscopic control to preparation of quasiequilibrium ensemble: we just prepare the
ensemble by macroscopic tools and then let them evolve due to natural dynamics (Liouville equation,
for example). Let us call this class {\it the primitive macroscopically definable ensembles}. These
ensembles appear as results (for $t>0$) of motions which start from the quasiequilibrium state (at
$t=0$). The main technical results of our work concern the computation of the manifold of primitive
macroscopically definable ensembles for a given system.

The importance of this class of ensembles is determined by the following hypothesis: {\bf The
hypothesis about the primitive macroscopically definable ensembles}. Any macroscopically definable
ensemble can be approximated by primitive macroscopically definable ensembles with appropriate
accuracy. In some limits we can attempt to say: ``with any accuracy". Moreover, this hypothesis with
``arbitrary small accuracy" can be found as the basic but implicit foundation of all nonequilibrium
kinetics theories which pretend to demonstrate a way from microdymamics to macrokinetics,  for
example in Zubarev nonequilibrium statistical operator theory \cite{22}, etc. This hypothesis allow
to describe nonequilibrium state as a result of evolution of quasiequilibrium state in time. Now we
have no other way to decribe the nonequilibrium state\footnote{There exists a series of papers with
discussion of Hamiltonian systems in so-called force thermostat, or, in particular, isokinetic
thermostat (see, for example, the review of D. Ruelle \cite{Ruelle}). These thermostats were invented
in computational molecular dynamics for acceleration of computations, as a technical trick. From
physical point of view this is a theory about a friction of particles on the space, the ``ether
friction" (the ``ether" is a theoretical substrate in the ancient physics). Of course, this theory is
mathematically consistent and perhaps it may be useful as the theory of special computations methods,
but a bridge between this theory and physics is desirable.}

The hypothesis about the primitive macroscopically definable ensembles is real hypothesis, it can
hold for different systems with different accuracy, it can be true or false. In some limits the set
of primitive macroscopically definable ensembles can be dense in the set of all macroscopically
definable ensembles, or can be not dense, etc. There is the significant difference between this
hypothesis and the {\it thesis} about macroscopically definable ensembles. The thesis can be
accepted, or not, the reasons for acceptance can be discussed, but nobody can prove or disprove the
definition, even the definition of the macroscopically definable ensembles.

\subsection{Irreversibility and initial conditions}

The choice of the initial state of the ensemble plays the crucial role in the thesis about the
macroscopically definable ensembles. The initial state is always taken as the quasiequilibrium
distribution which realizes the maximum of the entropy for given values of the macroscopic variables.
The choice of the initial state splits the time axis into two semi-axes: moving forward in time, and
moving backward in time, the observed non-order increases (the simplest example is the mixing of the
particles of different colors).

In some works, in order to achieve the ``true nonequilibrium'', that is, the irreversible motion
along the whole time axis, the quasiequilibrium initial condition is shifted into $-\infty$ in time.
This trick, however, casts some doubts, the major being  this:
 Most of the known equations of macroscopic dynamics which describe irreversible processes
have solutions which can be extended backwards in time only for finite times (or cannot be extended
at all). Such equations as the Boltzmann kinetic equation, diffusion equation, equations of chemical
kinetics and like do not allow for almost all their solutions to be extended backward in time for
indefinitely long. All motions have a ``beginning'' beyond which some physical properties of a
solution will be lost (often, positivity of distributions),
 although formally solutions may even exist, as in the case of chemical kinetics.

\subsection{Weak and strong tendency to equilibrium, shaking and short memory}

One aspect of irreversibility is the special choice of initial conditions. Roughly speaking, the
arrow of time is defined by the fact that the quasiequilibrium initial condition was in the past.

This remarkably simple observation does not, however, exhaust the problem of transition from the
reversible equations to irreversible macroscopic equations. One more aspect deserves a serious
consideration. Indeed, distribution functions tend to the equilibrium state  according to macroscopic
equations in a strong sense: deviations from the equilibrium tends to zero in the sense of most
relevant norms
 (in the $L^1$ sense, for example,
or even uniformly). On the contrast, for the Liouville equation, tendency to equilibrium ocures (if
at all) only in the weak sense: mean values of sufficiently ``regular'' functions on the phase space
do tend to their equilibrium values but the distribution function itself does not tend to the
equilibrium with respect to any norm, not even point-wise. This is especially easy to appreciate if
the initial state has been taken as the equipartition over some small bounded subset of the phase
space (the ``phase drop'' with small, but non-zero volume). This phase drop can mix over the phase
space, but for all the times it will remain  ``the drop of  oil in the water'', the density will
always be taking only two values, $0$ and $p>0$,
 and the volume of the set where the density is larger than zero will not be changing in time, of course.
So,  how to arrive from the weak convergence (in the sense of the convergence of the mean values), to
the strong convergence (to the $L^1$ or to the uniform convergence, for example)? In order to do
this, there are two basic constructions: The coarse-graining (shaking) in the sense of Ehrenfests',
and the short memory approximation.

The idea of coarse-graining dates back to P.\ and T.\ Ehrenfests, and it has been most clearly
expressed in their famous paper of 1911 \cite{Ehrenfest}. Ehrenfests considered a partition of the
phase space into small cells, and they have suggested to alter  the motions of the phase space
ensemble due to the Liouville equation with ``shaking'' - averaging of the density of the ensemble
over the phase cells. In the result of this process, the convergence to the equilibrium becomes
strong out of the weak. It is not difficult to recognize that ensembles with constant densities over
the phase cells are  quasiequilibria; corresponding macroscopic variables are integrals of the
density over the phase cells
 (``occupation numbers'' of the cells).
This generalizes to the following: alternations of the motion of the phase ensemble due to
microscopic equations with returns to the quasiequilibrium manifold, preserving the values of the
macroscopic variables. It is precisely this construction which serves for the point of departure for
many of the constructions below.

Another construction is the short memory approximation. The essence of it is the following: If one
excludes microscopic variables and assumes quasiequilibrium initial conditions, then it is possible
to derive integro-differential equations with retardation for the macroscopic variables (the way to
do this is not unique). The form of the resulting equations is approximately this:

\[ M(t)=\int_0^t K(t,t')[M(t')]dt', \]
where $K(t,t')$ is an operator (generally speaking, nonlinear) acting on $M(t')$. Once this equation
is obtained, one assumes that the kernels of these integro-differential equations decay at a
sufficiently high rate into the past (for example, exponentially, as $\|K(t,t')[M(t')]\|\le
\exp\{-(t-t')/ \tau \}\|M(t')\|$). This can be interpreted in the spirit of Ehrenfests': Every motion
which has begun sufficiently recently (the ``memory time'' $\tau$ before now) can be regarded as
being started from the quasiequilibrium. Thus, after each time $\tau$ has elapsed, the system can be
shaken in the sense of Ehrenfests - the result should not differ much.

\subsection{The essence of irreversibility in two words}

\noindent (i) The direction of the arrow of time is defined by the fact that only ``macroscopically
definable ensembles'' can be taken as initial conditions, that is, only quasiequilibrium ensembles
and what can be obtained from them when they are exposed to the true microscopic dynamics, or when
partial quasiequilibria are taken in positive time. {\it We} are created in such a way that we
prepare and control (in part)  the present, and observe what happens in the future. (In a sense, this
is a definition of the subjective time).

\noindent (ii) Microscopic dynamics can give only the weak convergence to the equilibrium,
convergence of mean values. Macroscopic variables tend to the equilibrium in the strong sense. The
passage from micro to macro occurs here with the help of Ehrenfests' coarse-graining procedure or its
analogs.

One might feel  uneasy  about the second of these points because the procedure of coarse-graining is
not the result of the equations of motion, and therefore it is somehow voluntary. The only hope to
lift this arbitrariness is that it  may well happen that, in the limit of a very large number of
particles, the perturbation caused by the coarse-graining can be made arbitrary small, for example,
by increasing the time interval between coarse-graining.

\subsection{Equivalence between trajectories and ensembles in the thermodynamic limit}
In the preceding sections we were speaking about the dynamics of ensembles. This apparently
contradicts the fact that the dynamics of a classical system goes along a single trajectory. Two
arguments make it possible to proceed from the trajectories to ensembles:

(i) High sensitivity of trajectories to external perturbations when the number of particles is large.
Arbitrary weak noise results in the stochastization of the motion.

(ii) In the thermodynamic limit, it is possible to partition the system into an arbitrary large
number of small but still macroscopic  sub-systems. Initial conditions in the sub-systems are
independent from one sub-system to another, and they cannot be assigned completely voluntary but are
taken from some distribution with  a fixed sum of mean values (an analog of the macroscopic
definability of ensembles). For spatially inhomogeneous systems, such small but still macroscopic
subsystems are defined in small and ``almost homogeneous'' volumes.

\subsection{Subjective time and irreversibility}
In our discussion, the source of the arrow of time is, after all, the asymmetry of the subjective
time of the experimentalist. {\it We prepare} initial conditions, and {\it after} that {\it we watch}
what will happen in the future but not what happened in the past. Thus, we obtain kinetic equations
for specifically prepared systems. How is this related to the dynamics of the real world? These
equations are applicable to real systems to the extent that the reality can be modeled with systems
with specifically prepared quasiequilibrium initial conditions. This is anyway less demanding than
the condition of  quasi-staticity of processes in classical thermodynamics. For this reason, versions
of nonequilibrium thermodynamics and kinetics based on this understanding of irreversibility allowed
to include such a variety of situations, and besides that, they include all classical equations of
nonequilibrium thermodynamics and kinetics.

\section{Geometrization of  irreversibility}\label{two}

\subsection{Quasiequilibrium manifold}

Let $E$ be a linear space, and $U\subset E$ be a convex subset, with a nonempty interior ${\rm int}
U$. Let a twice differentiable concave functional $S$ be  defined in ${\rm int} U$, and let $S$ is
continuous on $U$. According to the familiar interpretation, $S$ is the entropy, $E$ is an
appropriate space of distributions, $U$ is the cone of nonnegative distributions from $E$. Space $E$
is chosen in such a way that the entropy is well  defined on $U$.

Let $L$ be a closed linear subspace of space $E$, and $m:E\to E/L$ be the natural projection on the
factor-space. The factor-space $E/L$ will further play the role of the space of macroscopic variables
(in examples, the space of moments of the distribution).

\begin{figure}[p]
\begin{centering}
\includegraphics[width=150mm, height=180mm]{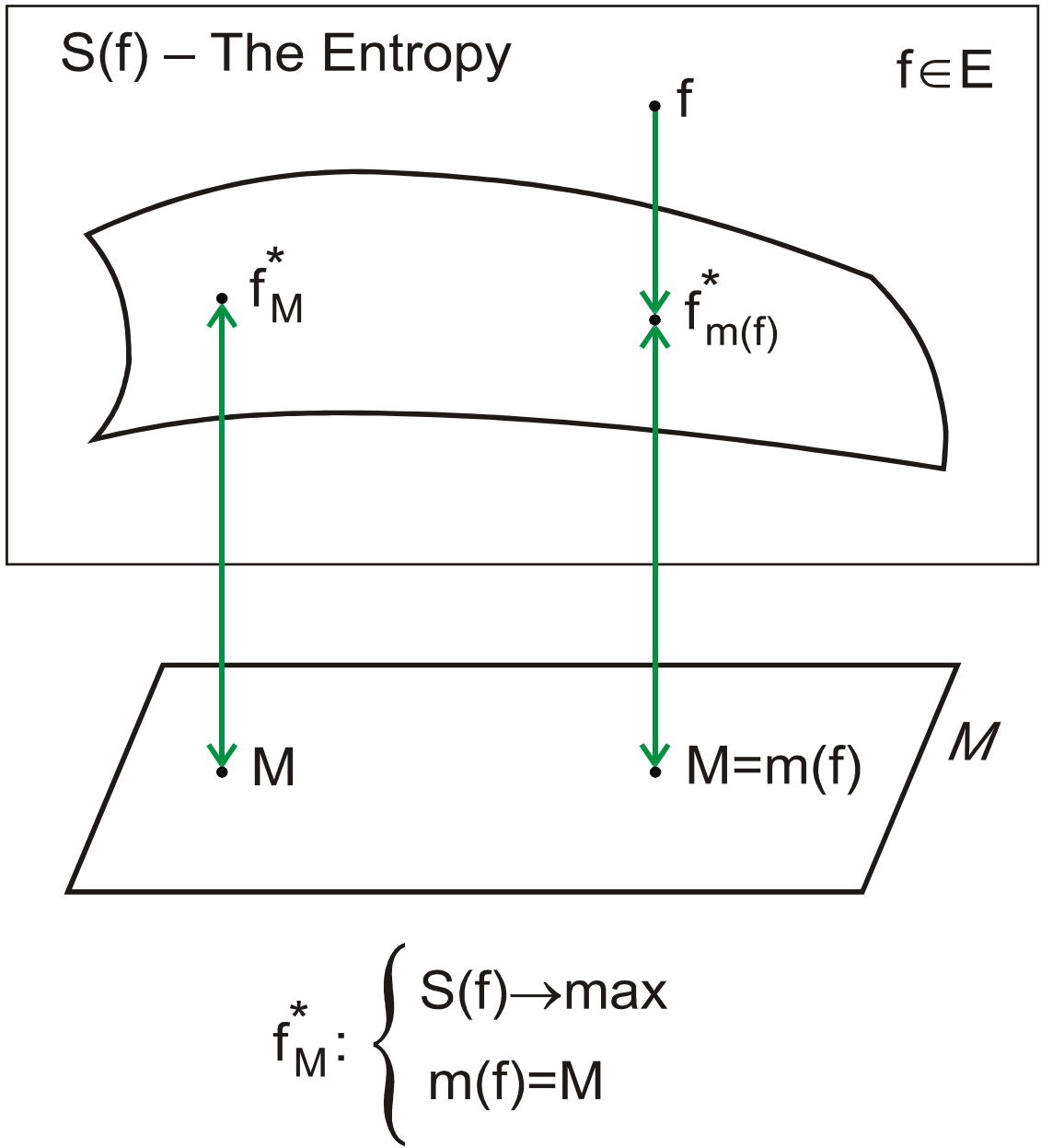}
\caption{Relations between a microscopic state $f$, the corresponding macroscopic state $M=m(f)$, and
quasiequilibria $f^*_M$.}
    \label{FigQEM}
\end{centering}
\end{figure}

For each $M\in {\rm int} U/L$ we define the quasiequilibrium, $f^*_M\in {\rm int }U$, as the solution
to the problem,
\begin{equation}
\label{QE2} S(f)\to{\rm max},\ m(f)=M.
\end{equation}
We assume that, for each $M\in{\rm int} U/L$, there exists the (unique) solution to the problem
(\ref{QE2}). This solution, $f^*_M$, is called the quasiequilibrium, corresponding to the value $M$
of the macroscopic variables. The set of quasiequilibria $f^*_M$ forms a manifold in ${\rm int} U$,
parameterized by the values of the macroscopic variables $M\in {\rm int} U/L$ (Fig. \ref{FigQEM}).

Let us specify some notations:  $E^T$ is the adjoint to the $E$ space.  Adjoint spaces and operators
will be indicated by $^T$, whereas notation $^*$ is earmarked for equilibria and quasiequilibria.

Furthermore, $[l,x]$ is the result of application of the functional $l\in E^T$ to the vector $x\in
E$.  We recall that, for an operator $A:E_1\to E_2$, the adjoint operator, $A^T:E_1^T\to E_2^T$ is
defined by the following relation:  For any $l\in E_2^T$ and $x\in E_1$,

\[ [l,Ax]=[A^Tl,x]. \]

Next, $D_fS(f)\in E^T$ is the differential of the functional $S(f)$, $D^2S(f)$ is the second
differential of the functional $S(f)$.  Corresponding quadratic functional $D^2S(f)(x,x)$ on $E$ is
defined by the Taylor formula,

\begin{equation}
\label{Taylor} S(f+x)=S(f)+[D_fS(f),x]+\frac{1}{2}D_f^2S(f)(x,x)+o(\|x\|^2).
\end{equation}
We keep the same notation for the corresponding symmetric bilinear form, $D_f^2S(f)(x,y)$, and also
 for the linear operator, $D_f^2S(f):E\to E^T$, defined by the formula,

\[ [D_f^2S(f)x,y]= D_f^2S(f)(x,y).\]

Here, on the left hand side there is the operator, on the right hand side there is the bilinear form.
Operator $D_f^2S(f)$ is symmetric on $E$, $D_f^2S(f)^T=D_f^2S(f)$.

Concavity of $S$ means that for any $x\in E$ the inequality holds, $D_f^2S(f)(x,x)\le 0$; in the
restriction onto the affine subspace parallel to $L$ we assume the strict concavity, $D_f^2S(f)(x,x)<
0$ if $x\in L$, and $x\ne0$.

A comment on the degree of rigor is in order: the statements which will be made below become theorems
or plausible hypotheses in specific situations. Moreover, specialization is always done with an
account for these statements
 in such a way as to simplify the proofs.

Let us compute the derivative $D_Mf^*_{M}$.  For this purpose, let us apply the method of Lagrange
multipliers:  There exists such a linear functional $\Lambda(M)\in (E/L)^T$, that

\begin{equation}
\label{Lagrange} D_fS(f)\big|_{f^*_{M}}=\Lambda(M)\cdot \mm,\ \mm(f^*_{M})=M,
\end{equation}

or

\begin{equation}
\label{Lagrange2} D_fS(f)\big|_{f^*_{M}}=\mm^T\cdot\Lambda(M),\ \mm(f^*_{M})=M.
\end{equation}
From equation (\ref{Lagrange2}) we get,

\begin{equation}
\mm(D_{M}f^*_{M})=1_{(E/L)},
\end{equation}
where we have indicated the space in which the unit operator is acting. Next, using the latter
expression, we transform the differential of the equation (\ref{Lagrange}),

\begin{equation}
D_{M}\Lambda=(\mm(D_f^2S)_{f^*_{M}}^{-1}\mm^T)^{-1},
\end{equation}
and, consequently,

\begin{equation}
\label{derivative} D_{M}f_{M}^*=(D_f^2S)_{f^*_{M}}^{-1}\mm^T(\mm(D_f^2S)_{f^*_{M}}^{-1}\mm^T)^{-1}.
\end{equation}
Notice that, elsewhere in equation (\ref{derivative}), operator $(D_f^2S)^{-1}$ acts on the linear
functionals from ${\rm im} \mm^T$. These functionals are precisely those which become zero on $L$
(that is, on ${\rm ker} \mm$), or, which is the same, those which can be represented as functionals
of macroscopic variables.

The tangent space to the quasiequilibrium manifold in the point $f_{M}^*$ is the image of the
operator $D_{M}f^{*}_{M}$:

\begin{equation}
\label{ann} {\rm im} \left(D_{M}f^{*}_{M}\right)=(D_f^2S)_{f^*_{M}}^{-1}{\rm im} \mm^T=
(D_f^2S)_{f^*_{M}}^{-1}{\rm Ann} L
\end{equation}
where ${\rm Ann} L$ (the annulator of $L$) is the set of linear functionals which become zero on $L$.
Another way to write equation (\ref{ann}) is the following:
\begin{equation}
\label{ann2} x\in {\rm im} \left(D_{M}f^{*}_{M}\right)\Leftrightarrow (D_f^2S)_{f^*_{M}}(x,y)=0,\
y\in L
\end{equation}
This means that $ {\rm im}\left(D_{M}f^{*}_{M}\right)$ is the orthogonal completement of $L$ in $E$
with respect to the scalar product,

\begin{equation}
\label{eproduct} \langle x|y\rangle_{f^*_{M}}=-(D_f^2S)_{f^*_{M}}(x,y).
\end{equation}

\begin{figure}[p]
\begin{center}
\begin{centering}
\includegraphics[width=150mm, height=190mm]{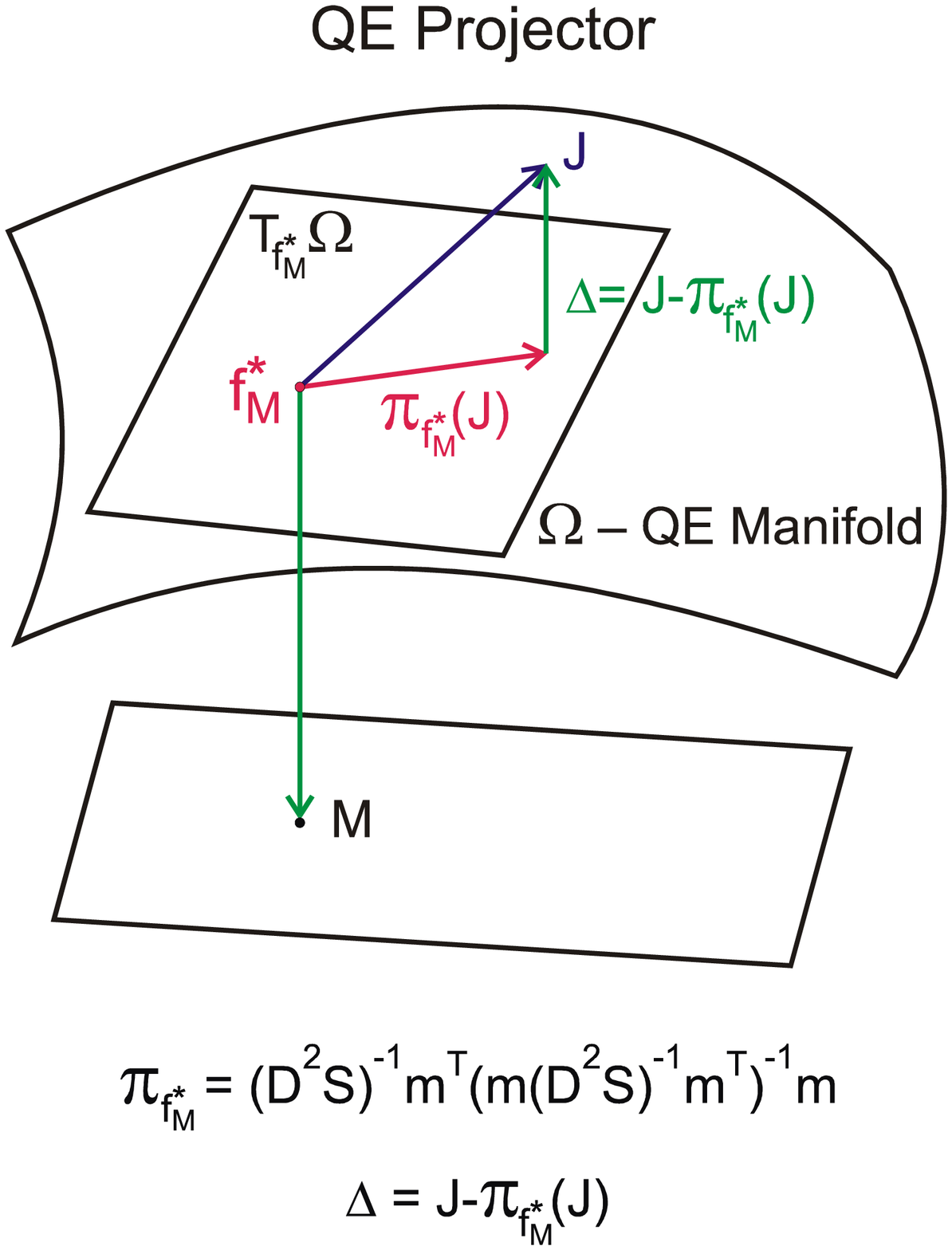}
\caption{Quasiequilibrium manifold $\Omega$, tangent space $T_{f^*_{M}}\Omega$, quasiequilibrium
projector $\pi_{f^*_{M}}$, and defect of invariance, $\Delta=\Delta_{f^*_{M}}=J-\pi_{f^*_{M}}(J)$.}
\label{QEpro}
\end{centering}
\end{center}
\end{figure}

The entropic scalar product (\ref{eproduct}) appears often in the constructions below. (Usually, this
becomes the scalar product indeed after the conservation laws are excluded). Let us denote as
$T_{f^*_{M}}={\rm im}(D_M f^*_{M})$ the tangent space to the quasiequilibrium manifold in the point
$f^*_M$. An important role in the construction of quasiequilibrium dynamics
 and its generalizations is played by the quasiequilibrium projector, an operator which projects
$E$ on $T_{f^*_{M}}$ parallel to $L$. This is the orthogonal projector with respect to the entropic
scalar product, $\pi_{f^*_{M}}:E\to T_{f^*_{M}}$:

\begin{equation}
\label{qeproj} \pi_{f^*_{M}}=\left(D_Mf^*_M\right)_M m=\left(D_f^2S \right)_{f^*_{M}}^{-1}m^T
\left(m\left(D_f^2S\right)_{f^*_{M}}^{-1}m^T\right)^{-1}m.
\end{equation}
It is straightforward to check the equality $\pi_{f^*_{M}}^2=\pi_{f^*_{M}}$, and the self-adjointness
of $\pi_{f^*_{M}}$ with respect to entropic scalar product (\ref{eproduct}). Thus, we have introduced
the basic constructions: quasiequilibrium manifold, entropic scalar product, and  quasiequilibrium
projector (Fig. \ref{QEpro}.

\subsection{Thermodynamic projector}\label{tp}

The construction of the quasiequilibrium allows for the following generalization: Almost every
manifold can be represented as a set of minimizers of the entropy under linear constraints. However,
in general, these linear constraints will depend on the point on the manifold.

So, let the manifold $\Omega=f_M\subset  U$ be given. This is a parametric set of distribution
function, however,  now macroscopic variables $M$ are not functionals on $R$ or $U$ but just
parameters defining the point on the manifold. The problem is how to extend definitions of $M$ onto a
neighborhood of $f_M$ in such a way that $f_M$ will appear as the solution to the variational
problem:

\begin{equation}
\label{smax2} S(f)\to\max,\ m(f)=M.
\end{equation}

For each point $f_M$, we identify $T_M\in E$, the tangent space to the manifold $\Omega$ in $f_M$,
and subspace $L_M\subset E$, which depends smoothly on $M$, and which has the property, $L_M\bigoplus
T_M=E$. Let us define $m(f)$ in the neighborhood of $f_M$ in such a way, that

\begin{equation}
\label{plane1} m(f)=M, \ {\rm if}\ f-f_M\in L_M.
\end{equation}

The point $f_M$ will be the solution of the quasiequilibrium problem (\ref{smax2}) if and only if
\begin{equation}
\label{plane2} D_fS(f)\big|_{f_M}\in {\rm Ann}\ L_M.
\end{equation}
That is, if and only if $L_M\subset {\rm ker} D_fS(f)\big|_{f_M}$. It is always possible to construct
subspaces $L_M$ with the properties just specified, at least locally, if the functional
$D_fS(f)\big|_{f_M}$ is not identically equal to zero on $T_M$.

The construction just described allows to consider practically any manifold as a quasiequilibrium.
This construction is required when one seeks the induced dynamics on a given manifold. Then the
vector fields are projected on $T_M$ parallel to $L_M$, and this preserves intact the basic
properties of the quasiequilibrium approximations.

It was proven \cite{InChLANL,CMIM} the {\bf theorem of uniqueness of the the thermodynamic
projector}:  There exists the unique operator which transforms the arbitrary vector field equipped
with the given Lyapunov function into a vector field with the same Lyapunov function (and also this
happens on any manifold which is not tangent to the level of the Lyapunov function).

Thermodynamic projector is constructed in the following way: Assume that the manifold $\Omega \subset
U$ is given, $f\in \Omega$ and $T$ is the tangent space to the manifold $\Omega$ in the point $f$.
Let us describe the construction of the thermodynamic projector onto tangent space $T$ in the point
$f$.

Let us consider $T_0$ that is a subspace of $T$ and which is annulled by the differential $S$ in the
point $f$:
\begin{equation}
T_0 = \{a\in T|(D_fS)(a)=0\}
\end{equation}

If $T_0 = T$, then the thermodynamic projector is the orthogonal projector on $T$ with respect to the
entropic scalar product $\langle |\rangle_f$ (\ref{eproduct}). Suppose that $T_0\neq T$. Let $e_g\in
T$, $e_g \perp T_0$ with respect to the entropic scalar product $\langle \mid \rangle_f$, and
$(D_fS)(e_g)=1$. These conditions define vector $e_g$ uniquely.

The projector onto $T$ is defined by the formula

\begin{equation}\label{ep}
P(J)=P_0(J)+e_g(D_fS)(J)
\end{equation}

\noindent where $P_0$ is the orthogonal projector with respect to the entropic scalar product
$\langle \mid \rangle_f$. For example, if $T$ a finite-dimensional space, then the projector
(\ref{ep}) is constructed in the following way. Let $e_1,..,e_n$ be a basis in $T$, and for
definiteness, $(D_fS)(e_1)\neq 0$.

\noindent 1) Let us construct a system of vectors

\begin{equation}
b_i=e_{i+1}-\lambda_i e_1, (i=1,..,n-1),
\end{equation}

\noindent where $\lambda_i=(D_fS)(e_{i+1})/(D_fS)(e_{1})$, and hence $(D_fS)(b_i)=0$. Thus,
$\{b_i\}_1^{n-1}$ is a basis in $T_0$.

\noindent2) Let us orthogonalize $\{b_i\}_1^{n-1}$ with respect to the entropic scalar product
$\langle \mid \rangle_f$ (\ref{eproduct}). We have got an orthonormal with respect to $\langle \mid
\rangle_f$ basis $\{g_i\}_1^{n-1}$ in $T_0$.

\noindent3) We find $e_g\in T$ from the conditions:

\begin{equation}
\langle e_g \mid g_i \rangle _f = 0, (i=1,..,n-1), (D_fS)(e_g)=1.
\end{equation}

\noindent and, finally we get

\begin{equation}\label{pfin}
P(J) = \sum_{i=1}^{n-1}g_i \langle g_i\mid J \rangle_f+e_g (D_fS)(J)
\end{equation}

If $(D_fS)(T)=0$, then the projector P is simply the orthogonal projector with respect to the
entropic scalar product. This is possible, for example, if $f$ is the global maximum of entropy point
(equilibrium). Then

\begin{equation}\label{peq}
P(J) = \sum_{i=1}^n{g_i\langle g_i|J \rangle_f}, \langle g_i|g_j \rangle = \delta_{ij}.
\end{equation}

If $(D_fS)(T)=0$ and $f$ is not equilibrium ($\Omega$ is tangent to the to the level of the entropy),
then the dynamic $\dot{f}=J(f)$ can be projected on $\Omega$ with preservation of dissipation only if
$(D_fS)(J(f))=0$ in this point.

\subsection{Quasiequilibrium approximation}

Let a kinetic equation be  defined in $U$:
\begin{equation}
\label{kin} \frac{df}{dt}=J(f).
\end{equation}
(This can be the Liouville equation, the Boltzmann equation, and so on, dependent on which level of
precision is taken for  the microscopic description.) One seeks the dynamics of the macroscopic
variables $M$. If we adopt the thesis that the solutions of the equation (\ref{kin}) of interest for
us begin on the quasiequilibrium manifold, and stay close to it for all the later times, then, as the
first approximation, we can take the quasiequilibrium approximation. It is constructed this way: We
regard $f$ as the quasiequilibrium, and write,
\begin{equation}
\label{qea1} \frac{dM}{dt}=m\left(J\left(f^*_M\right)\right).
\end{equation}
With this, the corresponding to $M$ point on the quasiequilibrium manifold moves according to the
following equation:
\begin{equation}
\label{qea2} \frac{df^*_{M(t)}}{dt}=(D_Mf^*_M)m(J(f^*_M))=\pi_{f^*_M}J(f^*_M),
\end{equation}
where $\pi_{f^*_M}$ is the quasiequilibrium projector (\ref{qeproj}). It is instructive to represent
solutions to equations of the quasiequilibrium approximation (\ref{qea2}) in the following way: Let
$T_{\tau}(f)$ be  the shift operator along the phase flow of equation (\ref{kin}) (that is,
$T_{\tau}(f)$ is solution to equation (\ref{kin}) at the time $t=\tau$ with the initial condition $f$
at $t=0$). Let us take the initial point $f_0=f^*_{M_0}$, and set, $f_{1/2}=T_{\tau}(f_0)$,
$M_1=m(f_{1/2})$, $f_1=f^*_{M_1}$, $\dots$, $f_{n+1/2}=T_{\tau}(f_n)$, $M_{n+1}=m(f_{n+1/2})$,
$\dots$. The sequence $f_n$ will be termed the {\it Ehrenfest's chain}. We set,
$f_{\tau}(n\tau)=f_n$. Then, $f_{\tau}(t)\to f(t)$, where $f(t)$ is the solution to the
quasiequilibrium approximation (\ref{qea2}), as $\tau\to0$, $n\to\infty$, $n\tau=t$.

\begin{figure}[p]
\begin{centering}
\includegraphics[width=150mm, height=190mm]{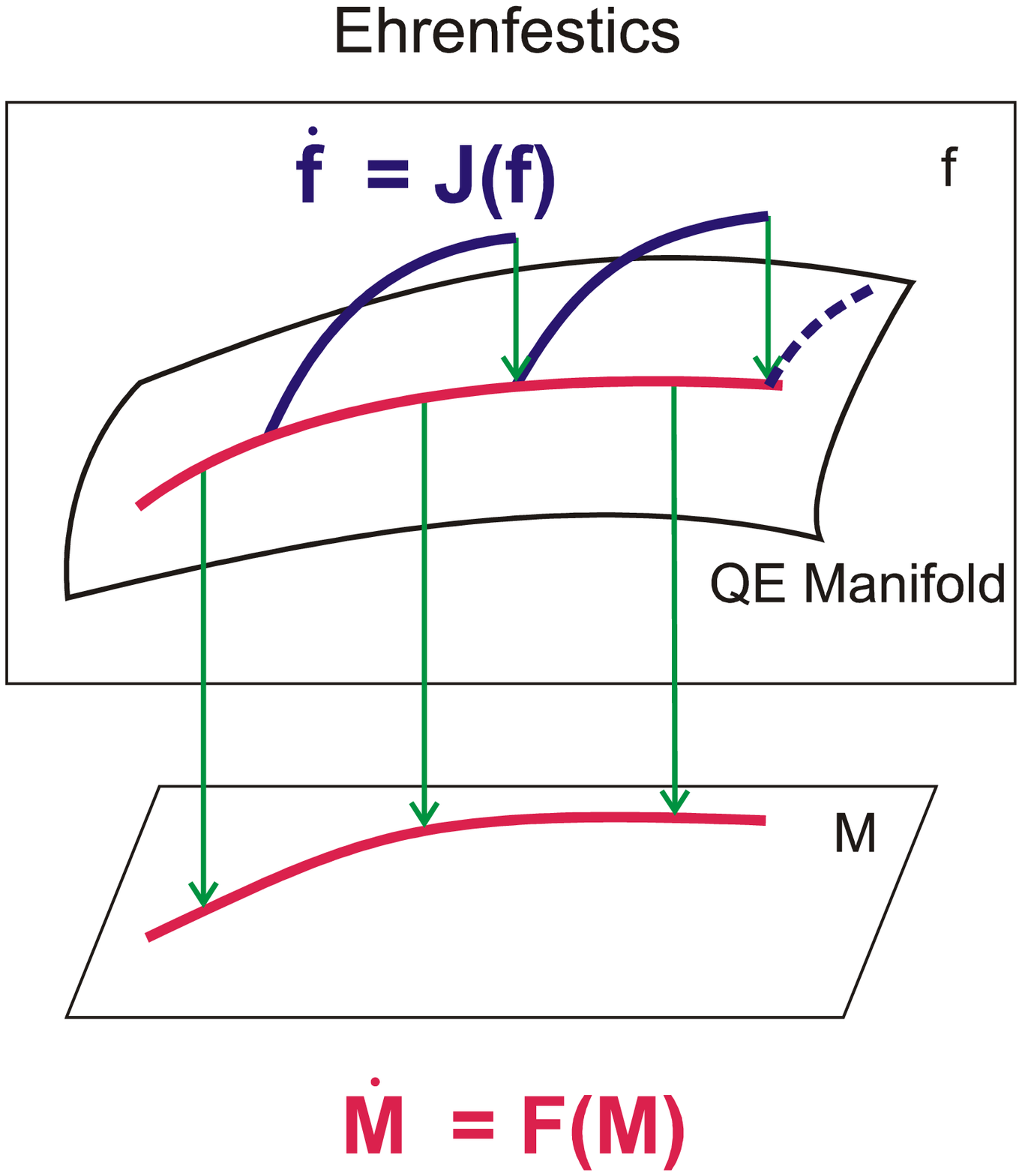}
\caption{Ehrenfest's chain over the quasiequilibrium manifold, and trajectory of the macroscopic
dynamics, $\dot{M}=F(M)$.} \label{FigEhr}
\end{centering}
\end{figure}

Let us notice that the way the entropy evolves in time according to the Ehrenfests' chain is defined
in the limit $\tau\to0$ solely by the way it evolves along trajectories of the kinetic equation
(\ref{kin}). Indeed, $f^*_M$ is the point of maximum of the entropy on the subspace defined by
equation, $m(f)=M$. Therefore, for
\[ S(f_{n+1/2})-S(f_{n+1})=o(\|f_{n+1/2}-f_{n+1}\|)=o(\tau),\]
it holds
\[ \sum_n| S(f_{n+1/2})-S(f_{n+1})|=o(n\tau)\to 0, \]
for $\tau\to0$, $n\to\infty$, $n\tau=const$. This simple observation has a rather important
implication: Let us denote as $dS(f)/dt$ the entropy production due to the original kinetic equation
(\ref{kin}), and as $(dS(f^*_M)/dt)_1$ its derivative due to the quasiequilibrium system
(\ref{qea2}). Then,
\begin{equation}
\label{equal1} (dS(f^*_M)/dt)_1=dS(f)/dt\big|_{f=f^*_M}.
\end{equation}
Let us give a different formulation of the latter identity. Let us term function $S(M)=S(f^*_M)$ {\it
the quasiequilibrium entropy}. Let us denote as $dS(M)/dt$ the derivative of the quasiequilibrium
entropy due to the quasiequilibrium approximation (\ref{qea1}). Then,
\begin{equation}
\label{equal2} \frac{dS(M)}{dt}=\frac{dS(f)}{dt}\bigg|_{f=f^*_M}.
\end{equation}
From the identity (\ref{equal1}), it follows {\bf the theorem about preservation of the type of
dynamics:}

(i) If, for the original kinetic equation (\ref{kin}),  $dS(f)/dt=0$ at $f=f^*_M$, then the entropy
is conserved due to the quasiequilibrium system (\ref{qea2}).

(ii) If, for the original kinetic equation (\ref{kin}),  $dS(f)/dt\ge0$ at  $f=f^*_M$, then, at the
same points, $f^*_M$, $dS(M)/dt\ge0$ due to the quasiequilibrium system (\ref{qea1}).

The theorem about the preservation of the type of dynamics\footnote{This is a rather old theorem, one
of us had published this theorem in 1984 already as textbook material (\cite{G1}, chapter 3
``Quasiequilibrium and entropy maximum", p. 37, see also the paper \cite{9}), but from time to time
different particular cases of this theorem are continued to be published as new results.}
demonstrates that, if there was no dissipation in the original system (\ref{kin}) (if the entropy was
conserved) then there is also no dissipation in the quasiequilibrium approximation. The passage to
the quasiequilibrium does not introduce irreversibility (the reverse may happen, for example, there
is no dissipation in the quasiequilibrium approximation for hydrodynamic variables as obtained from
the Boltzmann kinetic equation; though dissipation is present in the Boltzmann equation, it occurs in
different points but on the quasiequilibrium manifold of local Maxwellians the entropy production is
equal to zero). The same statement also hold for the thermodynamic projectors described in Section
\ref{tp}. On the other hand, the entropy production in the quasiequilibrium state is the same, as for
the quasiequilibrium system in the corresponding point, hence, if the initial system is dissipative,
then quasiequilibrium entropy production is nonnegative.

Usually, the original dynamics (\ref{kin}) does not leave  the quasiequilibrium manifold invariant,
that is, vector field $J(f)$ is not tangent to the quasiequilibrium manifold in all its points
$f^*_M$. In other words, the {\it condition of invariance},
\begin{equation}
\label{inv} (1-\pi_{f^*_M})J(f^*_M)=0,
\end{equation}
is not satisfied. The left hand side of the invariance condition (\ref{inv}) is of such an
outstanding importance that it deserves a separate name. We call it the {\it defect of invariance},
and denote it as $\Delta_{f^*_M}$. It is possible to consider the invariance condition as an
equation, and to compute corrections to the quasiequilibrium approximation $f^*_M$ in such a way as
to make it ``more invariant''. In those cases where the original equation (\ref{kin}) is already
dissipative, this route of corrections, supplemented by the construction of the thermodynamic
projector as in Section \ref{tp}, leads to an appropriate macroscopic kinetics \cite{8}.

However, here, we are mainly interested in the route ``from the very beginning'', from conservative
systems to dissipative. And here solving of the invariance equation does not help since it will lead
us to, while ``more invariant'', but still conservative dynamics. In all the approaches to this
problem (passage from the conservative to the dissipative systems), dissipation is introduced in a
more or less explicit fashion by various assumptions about the ``short memory''. The originating
point of our constructions will be the absolutely transparent and explicit approach of Ehrenfests.

\section{Natural projector and models of nonequilibrium dynamics}\label{natproj}

\subsection{Natural projector}\label{np}

\begin{figure}[p]
\begin{centering}
\includegraphics[width=150mm, height=210mm]{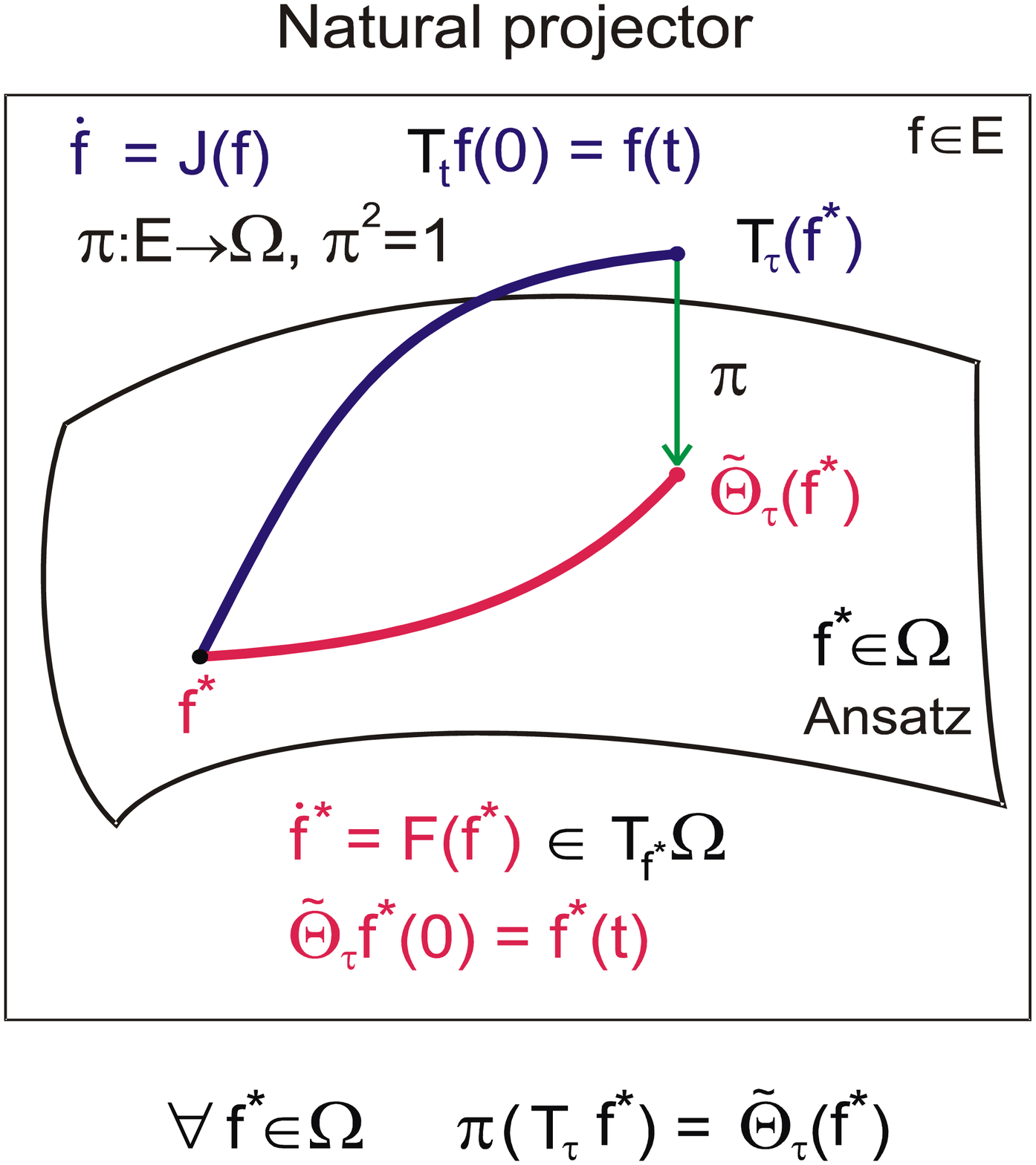}
\caption{Projection of segments of trajectories: The microscopic motion above the manifold $\Omega$
and the macroscopic motion on this manifold. If these motions began in the same point on  $\Omega$,
then, after time $\tau$, projection of the microscopic state onto  $\Omega$ should coincide with the
result of the macroscopic motion on $\Omega$. For quasiequilibrium  $\Omega$ projector $\pi: E
\rightarrow \Omega$ acts as $\pi(f)=f^*_{m(f)}$.} \label{FigNatPro}
\end{centering}
\end{figure}

So, let the original system (\ref{kin}) be conservative, and thus, $dS(f)/dt=0$. The idea of
Ehrenfests is to supplement the dynamics (\ref{kin}) by ``shakings''. Shakings are external
perturbations which are applied periodically with a fixed time interval $\tau$, and which lead to a
``forgetting'' of the small scale (nonequilibrium) details of the dynamics. For us here the shaking
is the replacement of $f$ with the quasiequilibrium distribution $f^*_{m(f)}$. In the particular case
which was originally considered in by Ehrenfests, the macroscopic variables $m(f)$ were the averages
of $f$ over cells in the phase space, while $f^*_{m(f)}$ was the cell-homogeneous distribution with
with the constant density within each cell equal to the corresponding cell-average of $f$. As we have
already mentioned it, in the limit $\tau\to0$, one gets back the quasiequilibrium approximation - and
the type of the dynamics is preserved. In this limit we obtain just the usual projection of the
vector field $J(f)$ (\ref{kin}) on the tangent bundle to the quasiequilibrium manifold. So, the
natural question appears: What will happen, if we will not just send $\tau$ to zero but will consider
finite, and even large, $\tau$? In such an approach, not just the vector fields are projected but
segments of trajectories. We shall term this way of projecting the {\it natural}. Let us now pose the
problem of the {\it natural projector} formally. Let $T_t(f)$ be the phase flow of the system
(\ref{kin}). We must derive a phase flow of the macroscopic system, $\Theta_t(M)$ (that is, the phase
flow of the macroscopic system, $dM/dt=F(M)$, which we are looking for), such that, for any $M$,
\begin{equation}
\label{match} m(T_{\tau}(f^*_M))=\Theta_{\tau}(M).
\end{equation}
That is, when moving along the macroscopic trajectory, after the time $\tau$ we must obtain the same
values of the macroscopic variables as if we were moving along the true microscopic trajectory for
the same time $\tau$, starting with the quasiequilibrium initial condition (Fig. \ref{FigNatPro}).

It is instructive to remark that, at finite $\tau$, the entropy growth follows immediately from
equation (\ref{match}) because $S(f)<S(f^*_{m(f)})$. The difference of the values of the entropy is
of the order $\|f-f^*_{m(f)}\|^2$, for the time $\tau$, thus, the first non-vanishing order in the
entropy production will be of the order of $\tau$. Let us find it.

We shall seek $F$ in terms of a series in $\tau$. Let us expand $F$ and {\it both} the sides of the
equation (\ref{match}) in powers of $\tau$ to second order, and find the expansion coefficients of
$F$ \cite{Lewis}\footnote{In this well known work \cite{Lewis} Lewis expanded only the right hand
side of equation (\ref{match}), and did not do the same also with the left hand side. There were some
hidden reason for this ``inconsistency": it was impossible to obtain the Boltzmann  equation without
such a deformation of expansion. We stress that our approach of matched expansion for exploring the
coarse-graining condition is, in fact, the exact (formal) statement that the unknown macroscopic
dynamics which causes the shift of $M$ on the left hand side of equation (\ref{match}) can be
reconstructed order-by-order to any degree of accuracy, whereas the low-order truncations may be
useful for certain physical situations. A thorough study of the cases beyond the lower-order
truncations is of great importance which is left for future work.}:
\begin{eqnarray*}
&&T_{\tau}(f^*_0)=f_0+df/dt\big|_{f_0}\tau+d^2f/dt^2\big|_{f_0}(\tau^2/2)+o(\tau^2),\\
&&\Theta_{\tau}(M_0)=M_0+dM/dt\big|_{M_0}\tau+d^2M/dt^2\big|_{M_0}(\tau^2/2)+o(\tau^2),\\
&&df/dt\big|_{f_0}=J(f_0),\ d^2f/dt^2\big|_{f_0}=D_fJ(f)\big|_{f_0}J(f_0),\\
&&dM/dt\big|_{M_0}=F(M_0),\ d^2M/dt^2\big|_{M_0}=D_MF(M)\big|_{M_0}F(M_0),\\ &&F(M)=F_0(M)+\tau
F_1(M)+o(\tau).
\end{eqnarray*}
Using these expansions in the condition for natural projector (\ref{match}), we get,
\begin{eqnarray*}
&&f_0=f^*_{M_0},\\ &&m(f_0)+\tau m(J(f_0))+(\tau^2/2)D_fJ(f)\big|_{f_0}J(f_0)+o(\tau^2)\\ &&=M_0+\tau
F_0(M_0)+\tau^2F_1(M_0)+(\tau^2/2)D_MF(M)\big|_{M_0}F(M_0)+o(\tau^2),
\end{eqnarray*}
whereupon,
\begin{eqnarray*}
&&F_0(M)=m(J(f^*_M)),\\
&&F_1(M)=(1/2)\left\{m(D_fJ(f)\big|_{f^*_M}J(f^*_M))-D_MF_0(M)\big|_{M}F_0(M)\right\}.
\end{eqnarray*}
Thus, the approximation $F_0$ is the quasiequilibrium, and using this fact in the expression for
$F_1$, after some transformation, we derive,
\begin{eqnarray*}
F_1&=&(1/2)\left\{m(D_fJ(f)\big|_{f^*_M}J(f^*_M))-D_M(m(J(f^*_M)))m(J(f^*_M))\right\}\\
&=&(1/2)\left\{m(D_fJ(f)\big|_{f^*_M}J(f^*_M))-m(D_fJ(f)\big|_{f^*_M}D_Mf^*_M)m(J(f^*_M))\right\}\\
&=&(1/2)m\left(D_fJ(f)\big|_{f^*_M}\left[J(f^*_M)-D_Mf^*_Mm(J(f^*_M))\right]\right)\\
&=&(1/2)m\left(D_fJ(f)\big|_{f^*_M}[1-\pi_{f^*_M}]J(f^*_M)\right)\\
&=&(1/2)m\left(D_fJ(f)\big|_{f^*_M}\Delta_{f^*_M}\right).
\end{eqnarray*}
Thus, the final form of the equation for the macroscopic variables $M$ may be written:
\begin{equation}
\label{macro2}
\frac{dM}{dt}=F(M)=m(J(f^*_M))+(\tau/2)m(D_fJ(f)\big|_{f^*_M}\Delta_{f^*_M})+o(\tau^2).
\end{equation}
It is remarkable the appearance of the defect of invariance in the second term (proportional to
$\tau$): If the quasiequilibrium manifold is invariant with respect to the microscopic dynamics, then
$F(M)$ is quasiequilibrium.

Let us compute the production of the quasiequilibrium entropy $S(M)=S(f^*_M)$ due to macroscopic
equations (\ref{macro2}), neglecting the higher-order term $o(\tau^2)$. $$
dS(f^*_M)/dt=(\tau/2)D_fS(f)\big|_{f^*_M}\pi_{f^*_M}D_fJ(f)\big|_{f^*_M}\Delta_{f^*_M}.  $$ We notice
that, $$ D_fS(f)\big|_{f^*_M}\pi_{f^*_M}=D_fS(f)\big|_{f^*_M}, $$ because $\pi_{f^*_M}$ is a
projector, and also because the thermodynamic condition  $${\rm ker}\ \pi_{f^*_M} \subset {\rm ker}\
D_fS(f)\big|_{f^*_M}$$ which follows from the definition of quasiequilibrium (\ref{QE2}). Next, by
our assumption, the system (\ref{kin}) conserves the entropy, $$dS(f)/dt=D_fS(f)\big|_fJ(f)=0. $$ Let
us differentiate the latter identity:
\begin{equation}
\label{id} D^2_fS(f)\big|_{f}J(f)+D_fS(f)\big|_fD_fJ(f)\big|_f=0.
\end{equation}
Thus, due to the right hand side of equation (\ref{macro2}),
\begin{eqnarray*}
\frac{dS(f^*_M)}{dt}&=&(\tau/2)D_fS(f)\big|_{f^*_M}D_fJ(f)\big|_{f^*_M}\Delta_{f^*_M}\\
&=&-(\tau/2)\left(D^2_fS(f)\big|_{f^*_M}J(f^*_M)\right)\Delta_{f^*_M}\\ &=&(\tau/2)\langle
J(f^*_M)|\Delta_{f^*_M}\rangle_{f^*_M},
\end{eqnarray*}
where we have used notation for entropic scalar product (\ref{eproduct}). Finally,
\[  \Delta_{f^*_M}=(1-\pi_{f^*_M})J(f^*_M)=(1-\pi_{f^*_M})^2J(f^*_M), \]
whereas projector $\pi_{f^*_M}$ is self-adjoint in the entropic scalar product (\ref{eproduct}).
Thus, $\langle J(f^*_M)|\Delta_{f^*_M}\rangle_{f^*_M}=\langle\Delta_{f^*_M}
|\Delta_{f^*_M}\rangle_{f^*_M}$, and
\begin{equation}
\label{production} \frac{dS(f^*_M)}{dt}=(\tau/2)\langle\Delta_{f^*_M} |\Delta_{f^*_M}\rangle_{f^*_M}.
\end{equation}

Thus, the quasiequilibrium entropy increases due to equation of macroscopic dynamics (\ref{macro2})
in those points of the quasiequilibrium manifold where the defect of invariance is not equal to zero.
This way we see how the problem of the natural projector (projected are not vector fields but
segments of trajectories) results in the dissipative equations. For specific examples see \cite{17}
where the second term in equation (\ref{macro2}) results in viscous terms in the Navier-Stokes
equations, diffusion and other dissipative contributions. However, it remains the undetermined
coefficient $\tau$. Formula (\ref{production}) gives the entropy production just proportional to the
time interval between subsequent coarse-graining. Of course, this could be true only for small enough
$\tau$, whereas we are mostly interested in the limit $\tau\to\infty$. It is only in this limit where
one can get rid of the arbitrariness in the choice of $\tau$ present in equations (\ref{macro2}) and
(\ref{production}). In order to do this, we need to study more carefully the structure of the
trajectories which begin on the quasiequilibrium manifold.

\subsection{One-dimensional model of nonequilibrium states}\label{plenka}

In the background of many derivations of nonequilibrium kinetic equations there is present the
following picture: Above each point of the quasiequilibrium manifold there is located a huge subspace
of nonequilibrium distributions with the same values of the macroscopic variables, as in the
quasiequilibrium point. It is as if the motion decomposes into two projections, above the point on
the quasiequilibrium manifold, and in the projection on this manifold. The motion in each layer above
the quasiequilibrium points  is extremely complicated, but fast, and everything quickly settles in
this fast motion.

However, upon a more careful looking into the motions which begin in the quasiequilibrium points, we
will observe that, above each point of the quasiequilibrium manifold it is located just a single
curve, and all the nonequilibrium (not-quasiequilibrium) states which come into the game form just a
one-dimensional manifold. It is namely this curve the construction of which we shall be dealing with
in this section.

This is the curve of {\it the primitive macroscopically definable ensembles}. These ensembles appear
as the result (for $t>0$) of motions which start from the quasiequilibrium state (at $t=0$).

For each value of the macroscopic variables $M$, and for each time $\tau$, we define $M_{-\tau}$ by
the following equality:

\begin{equation}
\label{back} m(T_{\tau}(f^*_{M_{-\tau}}))=M.
\end{equation}
In other words, $M_{-\tau}$ are those values of macroscopic variables which satisfy
$\Theta_{\tau}(M_{-\tau})=M$ for the natural projector (\ref{match}). Of course, it may well happen
that such $M_{-\tau}$ exists not for every pair $(M,\tau)$ but we shall assume here that for every
$M$ there exists such $\tau_M>0$ that there exists $M_{-\tau}$ for $0<\tau<\tau_M$.

A set of distributions, $q_{M,\tau}=T_{\tau}(f^*_{M_{-\tau}})$, forms precisely the desired curve of
nonequilibrium states with given values of $M$. Notice that, for each $\tau$, it holds,
$m(q_{M,\tau})=M$. The set $\{q_{M,\tau}\}$ for all possible $M$ and $\tau$ is positive invariant: If
the motion of the system starts on it at some time $t_0$, it stays on it also at $t>t_0$. If the
dependence $q_{M,\tau}$ is known, equations of motion in the coordinate system $(M,\tau)$ have a
simple form:
\begin{eqnarray}
\label{film1} \frac{d\tau}{dt}&=&1,\\\nonumber \frac{dM}{dt}&=&m(J(q_{M,\tau})).
\end{eqnarray}

The simplest way to study $q_{M,\tau}$ is through a consideration of a sequence of its derivatives
with respect to $\tau$ at fixed $M$. The first derivative is readily written as,
\begin{equation}
\label{dq} \frac{dq_{M,\tau}}{d\tau}\bigg|_{\tau=0}=J(f^*_M)-\pi_{f^*_M}J(f^*_M)=\Delta_{f^*_M}.
\end{equation}
By the construction of the quasiequilibrium manifold (we remind that $L={\rm ker}\ m$), for any $x\in
L$,

\[ S(f^*_M+\tau x)=S(f^*_M)-(\tau^2/2)\langle x|x\rangle_{f^*_M}+o(\tau^2).  \]
Therefore,

\[ S(q_{M,\tau})=S(f^*_M)-(\tau^2/2)\langle\Delta_{f^*_M} |\Delta_{f^*_M}\rangle_{f^*_M} +o(\tau^2). \]
Thus, to first order in $\tau$, we have, as expected.
\[ q_{M,\tau}=f^*_M+\tau\Delta_{f^*_M}+o(\tau). \]

Let us find $q_{M,\tau}$ to the accuracy of the order $o(\tau^2)$. To this end, we expand all the
functions in equation (\ref{back}) to the order of $o(\tau^2)$. With

\[ M_{-\tau}=M-\tau m(J(f^*_M))+\tau^2 B(M)+o(\tau^2), \]

where function $B$ is yet unknown, we write:

\[ f^*_{M_{-\tau}}=f^*_M-\tau D_Mf^*_M m(J(f^*_M))+\tau^2 D_Mf^*_M B(M)+(\tau^2/2)A_2(M)+o(\tau^2), \]
where
\begin{equation}
\label{A2} A_2(M)=\frac{d^2f^*_{M+tm(J(f^*_M))}}{dt^2}\bigg|_{t=0},
\end{equation}
and

\begin{eqnarray*}
T_{\tau}(x+\tau\alpha)&=&x+\tau\alpha+\tau J(x)+\tau^2 D_xJ(x)\big|_{x}\alpha\\
&&+(\tau^2/2)D_xJ(x)\big|_{x}J(x)+o(\tau^2),\\ T_{\tau}(f^*_{M_{-\tau}})&=&f^*_M-\tau
D_Mf^*_Mm(J(f^*_M))+\tau^2D_Mf^*_MB(M) +(\tau^2/2)A_2(M)\\&&+\tau
J(f^*_M)-\tau^2D_fJ(f)\big|_{f^*_M}D_Mf^*_M m(J(f^*_M))\\
&&+(\tau^2/2)D_fJ(f)\big|_{f^*_M}J(f^*_M)+o(\tau^2)\\
&=&f^*_M+\tau\Delta_{f^*_M}+(\tau^2/2)A_2(M)+(\tau^2/2)D_fJ(f)\big|_{f^*_M}(1-2\pi_{f^*_M})J(f^*_M)\\
&&+\tau^2 D_Mf^*_M B(M)+o(\tau^2).
\end{eqnarray*}
The latter somewhat lengthy expression simplifies significantly under the action of $m$. Indeed,
\begin{eqnarray*}
&& m(A_2(M))=d^2[M+tm(J(f^*_M))]/dt^2=0,\\ && m(1-\pi_{f^*_M})=0,\\ && m(D_Mf^*_M)=1.
\end{eqnarray*}
Thus,
\[
m(T_{\tau}(f^*_{M_{-\tau}}))=M+(\tau^2/2)m(D_fJ(f)\big|_{f^*_M}(1-2\pi_{f^*_M})J(f^*_M))+\tau^2B(M)+o(\tau^2),
\]
\[ B(M)=(1/2)m(D_fJ(f)\big|_{f^*_M}(2\pi_{f^*_M}-1)J(f^*_M)). \]
Accordingly, to second order in $\tau$,
\begin{eqnarray}
\label{film2} q_{M,\tau}&=&T_{\tau}(f^*_{M_{-\tau}})\\\nonumber
&=&f^*_M+\tau\Delta_{f^*_M}+(\tau^2/2)A_2(M) \\\nonumber
&&+(\tau^2/2)(1-\pi_{f^*_M})D_fJ(f)\big|_{f^*_M}(1-2\pi_{f^*_M})J(f^*_M)+o(\tau^2).
\end{eqnarray}
Notice that, besides the dynamic contribution of the order of $\tau^2$ (the last term), there appears
also the term $A_2$ (\ref{A2}) which is related to the curvature of the quasiequilibrium manifold
along the quasiequilibrium trajectory.

Let us address the behavior of the entropy production in the neighborhood of $f^*_M$. Let $x\in L$
(that is, $m(x)=0$). The production of the quasiequilibrium entropy, $\sigma^*_M(x)$, equals, by
definition,
\begin{equation}
\label{production2} \sigma_M^*(x)=D_MS(f^*_M)\cdot m(J(f^*_M+x)).
\end{equation}
Equation (\ref{production2}) gives the rate of entropy change under the motion of the projection of
the state onto the quasiequilibrium manifold if the true trajectory goes through the point $f^*_M+x$.
In order to compute the right hand side of equation (\ref{production2}), we use essentially the same
argument, as in the proof of the entropy production formula (\ref{production}). Namely,  in the point
$f^*_M$, we have $L\subset {\rm ker} D_fS(f)\big|_{f^*_M}$, and thus
$D_fS(f)\big|_{f^*_M}\pi_{f^*_M}= D_fS(f)\big|_{f^*_M}$. Using this, and the fact that entropy
production in the quasiequilibrium approximation is equal to zero, equation (\ref{production2}) may
be written,
\begin{equation}
\label{production3} \sigma_M^*(x)=D_fS(f)\big|_{f^*_M}(J(f^*_M+x)-J(f^*_M)).
\end{equation}
To the linear order in $x$, the latter expression reads:
\begin{equation}
\label{production4} \sigma_M^*(x)=D_fS(f)\big|_{f^*_M}D_fJ(f)\big|_{f^*_M}x.
\end{equation}
Using the identity (\ref{id}), we obtain in equation (\ref{production4}),
\begin{equation}
\label{production5} \sigma_M^*(x)=-D_f^2S(f)\big|_{f^*_M}(J(f^*_M),x) =\langle
J(f^*_M)|x\rangle_{f^*_M}.
\end{equation}
Because $x\in L$, we have $(1-\pi_{f^*_M})x=x$, and
\begin{eqnarray*}
\langle J(f^*_M)|x\rangle_{f^*_M}&=&\langle J(f^*_M)|(1-\pi_{f^*_M})x\rangle_{f^*_M}\\&=& \langle
(1-\pi_{f^*_M})J(f^*_M)|x\rangle_{f^*_M}=\langle \Delta_{f^*_M}|x\rangle_{f^*_M}.
\end{eqnarray*}
Thus, finally, the entropy production in the formalism developed here, to the linear order reads,
\begin{equation}\label{sigma}
\sigma_M^*(x)=\langle \Delta_{f^*_M}|x\rangle_{f^*_M}.
\end{equation}

\subsection{Stability of quasiequilibrium manifolds}\label{stability}

The notion of stability does not cause essential difficulties  when it goes about an invariant
manifold, it is stable if, for any $\epsilon>0$, there exist such $\delta>0$ that a motion which has
started at $t=0$ at the distance (in some appropriate sense) less than $\delta$ from the manifold
will not go away further than $\epsilon$ at any $t>0$.

However, this  is not so for a non-invariant manifold, and, probably, it is not possible to give a
useful for all the cases formalization of the notion of {\it stability of the quasiequilibrium
manifold}, in the spirit of motions going not far away when  started sufficiently close to the
manifold (indeed, what is here ``sufficiently close'' and ``not far''?). In spite of that, expression
(\ref{film2}) gives important opportunity to measure the stability. Indeed, let us consider how the
entropy production depends on $\tau$, that is, let us study the function,
\begin{equation}
\label{sigmatau} \sigma_M(\tau)=\langle \Delta_{f^*_M}|q_{M,\tau}\rangle_{f^*_M}.
\end{equation}
It is natural to expect that $\sigma_M(\tau)$ initially increases, and then it saturates to some
limiting value.
The question is, however, how function $\sigma_M(\tau)$ behaves at $t=0$, is it concave or is it
convex in this point? If  function  $\sigma_M(\tau)$ is concave,
$d^2\sigma_M(\tau)/d\tau^2\big|_{\tau=0}<0$, then the speed with which it grows reduces immediately,
and one can even estimate the limiting value,
\[ \sigma_M^*=\lim_{\tau\to\infty}\sigma_M(\tau), \]
using the first Pad\'e approximate:
\begin{eqnarray}
\label{Pade} &&\sigma_M(\tau)=a\tau/(1+b\tau)=a\tau-ab\tau^2+\dots\\\nonumber
&&\sigma_M^*=a/b=-\frac{2(d\sigma_M(\tau)/d\tau\big|_{\tau=0})^2}{d^2\sigma_M(\tau)/d\tau^2\big|_{\tau=0}}.
\end{eqnarray}
Concavity of  $\sigma_M(\tau)$ at $\tau=0$ ($d^2\sigma_M(\tau)/d\tau^2\big|_{\tau=0}<0$) is analogous
to a soft instability: The motion does not run too far away, and it is possible to estimate where it
will stop, see equation (\ref{Pade}). However, if $d^2\sigma_M(\tau)/d\tau^2\big|_{\tau=0}>0$, then
this is analogous to a  hard instability, and none of the estimates like (\ref{Pade}) work. Thus,
everything is defined by the sign of the scalar product,
\begin{equation}
\label{sign} \frac{d^2\sigma_M(\tau)}{d\tau^2}\bigg|_{\tau=0}=\langle \Delta_{f^*_M}|A_2(M)+
D_fJ(f)\big|_{f^*_M}(1-2\pi_{f^*_M})J(f^*_M)\rangle_{f^*_M}.
\end{equation}
If this expression is negative, then the Pad\'e estimate (\ref{Pade}) gives:
\begin{equation}
\label{estimate} \sigma_M^*=-\frac{2\langle\Delta_{f^*_M}|\Delta_{f^*_M}\rangle^2_{f^*_M}} {\langle
\Delta_{f^*_M}|A_2(M)+ D_fJ(f)\big|_{f^*_M}(1-2\pi_{f^*_M})J(f^*_M)\rangle_{f^*_M}}.
\end{equation}
In the opposite case, if the sign of the expression (\ref{sign}) is {\it positive}, we call the
quasiequilibrium manifold {\it unstable}.

Equation (\ref{estimate}) allows us to estimate the parameter $\tau$ in the equations of the method
of natural projector. To this end, we make use of equation (\ref{production}):
\[ (\tau/2)\langle\Delta_{f^*_M}|\Delta_{f^*_M}\rangle_{f^*_M}=\sigma^*_{M}, \]
whereupon,
\begin{equation}
\label{tau} \tau\approx-\frac{4\langle\Delta_{f^*_M}|\Delta_{f^*_M}\rangle_{f^*_M}} {\langle
\Delta_{f^*_M}|A_2(M)+ D_fJ(f)\big|_{f^*_M}(1-2\pi_{f^*_M})J(f^*_M)\rangle_{f^*_M}},
\end{equation}
if the denominator assumes negative values. In this case, there are no free parameters left in
equation (\ref{macro2}).

Above, the parameter $\tau$, or the time of ``leaving the initial quasiequilibrium condition'', has
been appearing explicitly in the equations. Except for the case of linear quasiequilibrium manifolds
where the formal limit $\tau\to\infty$ can be addressed to derive generalized fluctuation-dissipation
relations \cite{GK02}, this may be not the best way to do in the general, nonlinear case.

\subsection{Curvature and entropy production:  Entropic circle and first kinetic equations}

In a consequent geometric approach to the problem of constructing the one-dimensional model of
nonequilibrium states it is sufficient to consider the entropic parameter, $\delta S=S^*(M)-S$.
Within this parameterization of the one-dimensional curve of the nonequilibrium states, one has to
address functions $\sigma_M(\Delta S)$, rather than $\sigma_M(\tau)$ (\ref{sigmatau}), whereas their
Pad\`e approximates can be constructed, in turn, from expansions in $\tau$.

In order to give an example here, we notice that the simplest geometric estimate amounts to
approximating the trajectory $q_{M,\tau}$ with a second order curve. Given $\dot{q}_{M,\tau}$ and
$\ddot{q}_{M,\tau}$ (\ref{film2}), we construct a tangent circle (in the entropic metrics,
$\langle|\rangle_{f^*_M}$, since the entropy is the integral of motion of the original equations).
For the radius of this circle we get,
\begin{equation}
\label{radius}
R=\frac{\langle\dot{q}_{M,0}|\dot{q}_{M,0}\rangle_{f^*_M}}{\sqrt{\langle\ddot{q}_{\perp\
M,0}|\ddot{q}_{\perp\ M,0}\rangle_{f^*_M}}},
\end{equation}
where
\begin{eqnarray*}
\dot{q}_{M,0}&=&\Delta_{f^*_M},\\ \ddot{q}_{\perp\ M,0}&=&\ddot{q}_{M,0}-\frac{\langle
\ddot{q}_{M,0}|\Delta_{f^*_M}\rangle_{f^*_M}\Delta_{f^*_M}}
{\langle\Delta_{f^*_M}|\Delta_{f^*_M}\rangle_{f^*_M}},\\
\ddot{q}_{M,0}&=&(1-\pi_{f^*_M})D_fJ(f)\big|_{f^*_M}(1-2\pi_{f^*_M})J(f^*_M)
+\left(D_M\pi_{f^*_M}\right)m(J(f^*_M)).
\end{eqnarray*}

Let us represent the microscopic motion as a circular motion along this entropic circle with constant
velocity $\dot{q}_{M,0}=\Delta_{f^*_M}$. When the microscopic motion passed the quarter of the
circle, the entropy production started to decrease and it became zero after the halve of the circle.
Hence, after passing the quarter of the circle, this model should be changed. The time of the motion
along the quarter of the model entropic circle is:
\begin{equation}\label{tauR}
\tau \approx \frac{\pi}{2}\sqrt{ \frac{\langle\Delta_{f^*_M}|\Delta_{f^*_M}\rangle_{f^*_M}}
{\langle\ddot{q}_{\perp\ M,0}|\ddot{q}_{\perp\ M,0}\rangle_{f^*_M}}}.
\end{equation}

After averaging on the $1/4$ of this circle circular motion we obtain the macroscopic
equations\footnote{This averaging makes sense for {\it conservative} microdynamics, and for
dissipative microdynamics the model of uniform circular motion along the entropic circle should be
improved by taking into account the acceleration along the circle.}
\begin{eqnarray}\label{firsteq}
&&{dM \over dt} = m\left(J\left(f^*_M + {2 \over \pi} R {\Delta_{f^*_M} \over \|\Delta_{f^*_M}\|} +
\left(1- {2 \over \pi}\right) R {\ddot{q}_{\perp\ M,0} \over \| \ddot{q}_{\perp\
M,0}\|}\right)\right) = m(J(f^*_M)) + \\ && + {2 \over \pi} {R \over \|\Delta_{f^*_M}\|}
m\left(D_fJ(f)\big|_{f^*_M}(\Delta_{f^*_M})\right) + \left(1- {2 \over \pi}\right) {R \over \|
\ddot{q}_{\perp\ M,0}\|}m\left(D_fJ(f)\big|_{f^*_M}(\ddot{q}_{\perp\ M,0}) \right) + o(R). \nonumber
\end{eqnarray}
\noindent where $\|y\|=\sqrt{\langle y | y \rangle_{f^*_M}}$.

Equations (\ref{firsteq}) contain no undetermined parameters. This is the simplest example of the
general macroscopic equations obtained by the natural projector. The coefficients ($2/ \pi$, etc.)
can be corrected, but the form is more universal. The entropy production for equations
(\ref{firsteq}) is proportional both to the defect of invariance and to the radius of curvature:
\begin{equation}\label{firstpro}
\sigma_M={2 \over \pi} R  \|\Delta_{f^*_M}\|.
\end{equation}
This equation demonstrates the thermodynamical sense of curvature of the curve of nonequilibrium
states. The combination ${\mbox{defect of invariance} \over \mbox{curvature}}$ is the dissipation.
(It should be remained that all the scalar products and norms are {\it entropic}).


\section{The film of non-equilibrium states}\label{4}

\subsection{Equations for the film}\label{4.1}

The set $q_{M,\tau}$ in a space $E$ generates a surface parameterized by ``two variables": A scalar,
$\tau\geq 0,$ and value of macroscopic variables, $M,$ with condition
\begin{eqnarray}
M=m(q_{M,\tau}).\label{47}
\end{eqnarray}
We call this surface {\it the film of non-equilibrium states} or simply {\it the film.} It consists
of {\it the primitive macroscopically definable ensembles}, the results (for $t>0$) of motions which
start from the quasiequilibrium state (at $t=0$).

\begin{figure}[p]
\begin{centering}
\includegraphics[width=150mm, height=160mm]{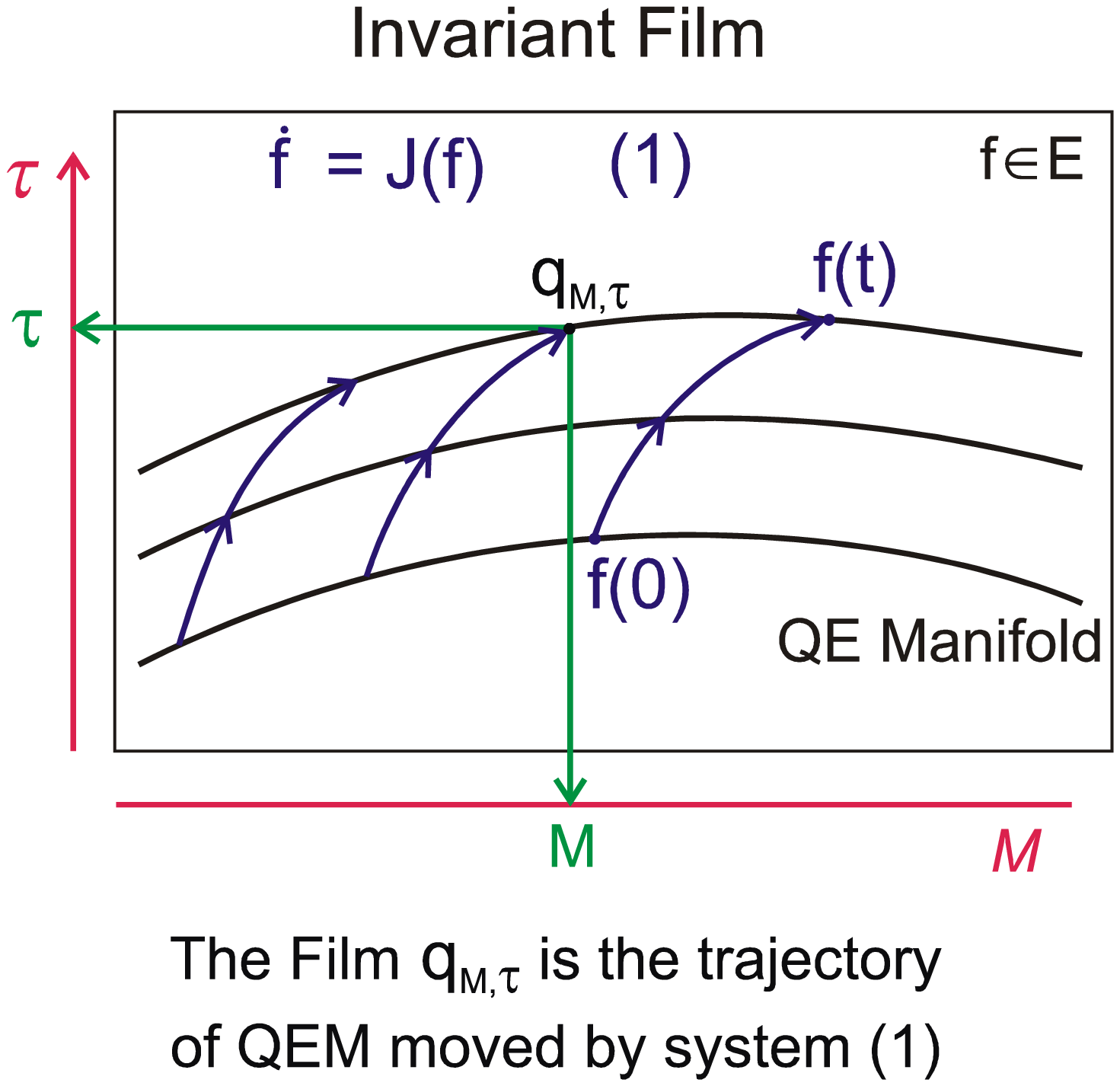}
\caption{The film of nonequilibrium states as the trajectory of motion of the quasiequilibrium
manifold due to microscopic dynamics.} \label{FigFilm}
\end{centering}
\end{figure}

For each $\tau \geq 0$ {\it the section of the film} is defined: the set, $q_{M,\tau},$ for given
$\tau.$ It is parameterized by the value of $M.$ For $\tau=0$ the section of the film coincides with
the quasiequilibrium manifold. The film itself can be considered as a trajectory of motion of the
section under variation of $\tau \in [0;+\infty)$ (Fig. \ref{FigFilm}). It is not difficult to write
down equations of this motion using definition of $q_{M,\tau}:$
\begin{equation}
q_{M,\tau}=T_{\tau}f^*_{M_{-\tau}},\label{48}
\end{equation}
where $T_{\tau}$ is shift in time in accordance with the original dynamical system, $M_{-\tau}$ is
defined with equation (\ref{back}).

For small $\Delta \tau$
\begin{eqnarray}
q_{M,\tau+\Delta\tau}=q_{M-\Delta M,\tau}+J(q_{M,\tau})\Delta \tau+o(\Delta \tau),\label{49}
\end{eqnarray}
where $\Delta M=mJ(q_{M,\tau})\Delta \tau.$ Hence,
\begin{eqnarray}
{dq_{M,\tau}\over dt}=(1-D_{M}q_{M,\tau}m)J(q_{M,\tau}).\label{50}
\end{eqnarray}
Initial condition for equation (\ref{50}) is the quasiequilibrium
\begin{eqnarray}
q_{M,0}=f^*_M.\label{51}
\end{eqnarray}

Equation (\ref{50}) under initial condition (\ref{51}) defines the film of non-equilibrium states in
the space $E.$ This film is a minimal positive invariant set (i.e invariant with respect to shift
$T_{\tau}$ by positive times $\tau >0$), including the quasiequilibrium manifold, $f^*_M.$ The
kinetics we are interested in occurs only on this film.

Investigation of non-equilibrium kinetics can be separated into two problems:

\noindent i) Construction of the film of non-equilibrium states: solution of equation (\ref{50})
under initial conditions (\ref{51}).

\noindent ii) Investigation of the motion of the system on the film.

Naturally, it should be assumed that the film will be constructed only approximately. Therefore, the
second problem should be separated in two again:

\noindent ii1) Construction of projection of initial vector field, $J,$ on the approximately found
film, and construction of equations for $M$ and $\tau.$

\noindent ii2) Investigation and solution of equations for $M$ and $\tau.$

It should be especially emphasized that existence of the film is not significantly questionable
(though, of course, the deriving of the theorems of existence and uniqueness for (\ref{50}),
(\ref{51}) can turn into a complicated mathematical problem). In contrast, existence of kinetic
coefficients (viscosity etc.), and generally, the fast convergence of $dM/dt$ to a certain dependence
$dM/dt(M)$ is essentially a hypothesis which is not always true.

Below we will be solving, mainly, the problem of construction of equations: problems ii1) and ii2).
And we will begin with the problem ii2). Thus, let the film be approximately constructed.

\subsection{Thermodynamic projector on the film}\label{4.2}

We need the projector in order to project the vector field on the tangent space. The idea of
thermodynamic projector \cite{3} consists of a description of every manifold (subject to certain
requirements of transversality) as the quasiequilibrium one. For this, one constructs a projection of
a neighborhood of the manifold on it, and later, the required projector is obtained by linearization.

The projection of the neighborhood on the manifold should satisfy essentially only one condition: a
point of manifold must be the point of maximum of the entropy on its preimage. If the preimage of
point $f^*$ is a domain in the affine subspace, $L_{f^*}\subset E,$ then required condition is:
\begin{eqnarray}
D_fS^*(L_{f^*}-f^*)\equiv 0.\label{52}
\end{eqnarray}
where $L_{f^*}-f^*$ is already the linear subspace in $E$.

For such projections, a dissipative vector field is projected into a dissipative one, and a
conservative vector field (with the entropy conservation) is projected into a conservative one, i.e.
the entropy balance is exact. Thus, let the film, $q_{M,\tau},$ be defined. Let us construct for it a
thermodynamic projector.

\begin{figure}[p]
\begin{centering}
\includegraphics[width=150mm, height=155mm]{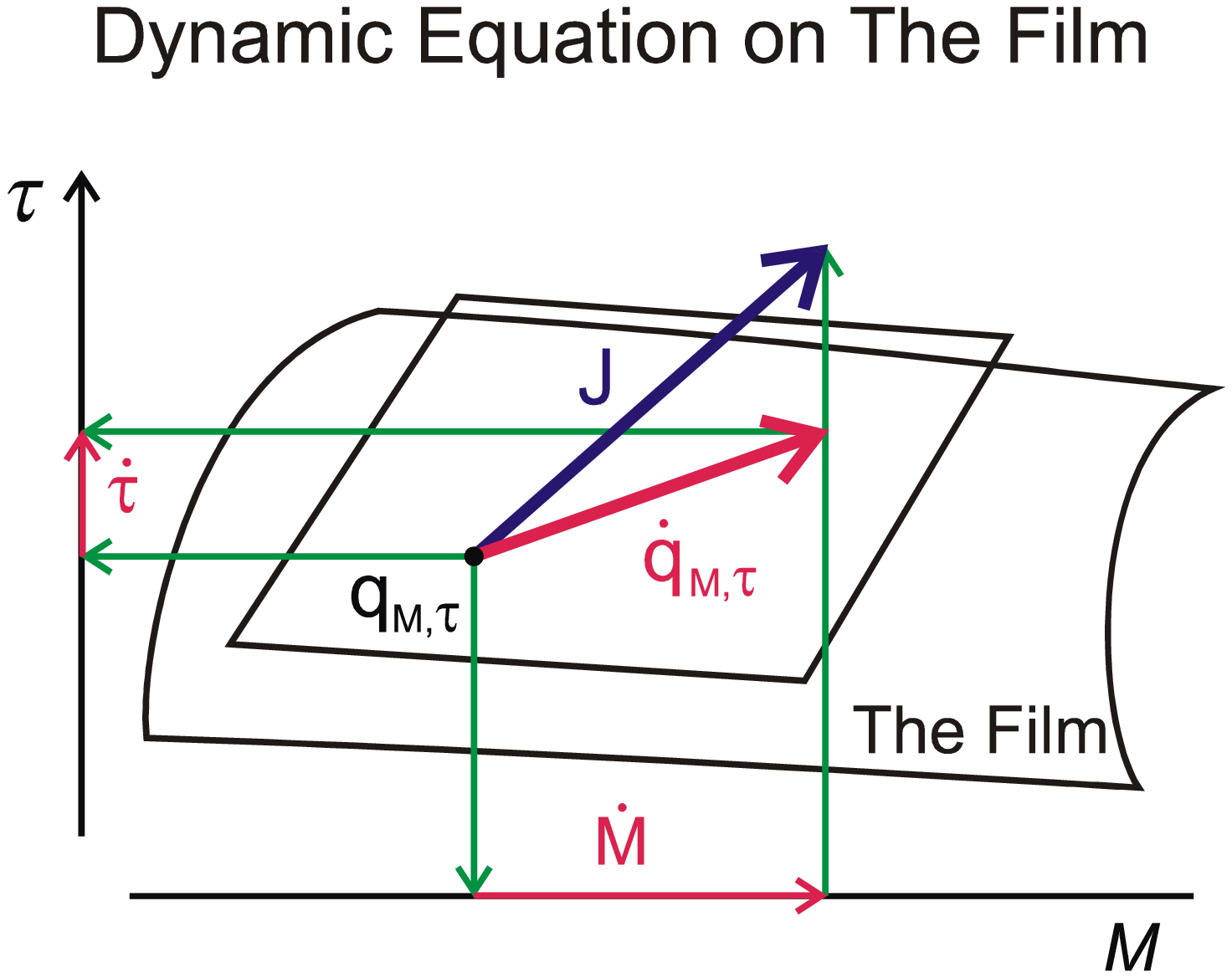}
\caption{Dynamics on the film: $\dot{M} = mJ(q_{M,\tau}),$ $\dot{\tau}= -{D_fS|_{q_{M,\tau}}
D_Mq_{M,\tau} \dot M\over D_fS|_{q_{M,\tau}}D_{\tau}q_{M,\tau}}$.} \label{FigDinFilm}
\end{centering}
\end{figure}

Under small variation of variables $M$ and $\tau$
\begin{eqnarray}
\Delta q_{M,\tau}&=&D_Mq_{M,\tau}\Delta M+D_{\tau}q_{M,\tau}\Delta \tau+o(\Delta M,\Delta
\tau),\nonumber\\* \Delta S&=&D_fS\left|_{q_{M,\tau}}\right.\Delta q_{M,\tau}+o(\Delta M,\Delta
\tau).\label{53}
\end{eqnarray}
After simple transformations we obtain:
\begin{eqnarray}
\Delta \tau&=&{1\over D_fS|_{q_{M,\tau}}D_{\tau}q_{M,\tau}}+o(\Delta M,\Delta S ),\nonumber\\* \Delta
q_{M,\tau}&=&\left[1-{D_{\tau}q_{M,\tau}D_fS|_{q_{M,\tau}}\over D_fS|_{q_{M,\tau}}D_{\tau}q_{M,\tau}
}\right]D_Mq_{M,\tau}\Delta M\nonumber\\*&+&{1\over
D_fS|_{q_{M,\tau}}D_{\tau}q_{M,\tau}}D_{\tau}q_{M,\tau}\Delta S+o(\Delta M,\Delta S).\label{54}
\end{eqnarray}
From this formulae we obtain thermodynamic projector for $J$, $\pi_{\rm td}$:
\begin{eqnarray}
\pi_{\rm td}|_{q_{M,\tau}}J=\left[1-{D_{\tau}q_{M,\tau}D_fS|_{q_{M,\tau}}\over
D_fS|_{q_{M,\tau}}D_{\tau}q_{M,\tau}
}\right]D_Mq_{M,\tau}mJ+{D_{\tau}q_{M,\tau}D_fS|_{q_{M,\tau}}\over
D_fS|_{q_{M,\tau}}D_{\tau}q_{M,\tau} }J.\label{55}
\end{eqnarray}
For conservative systems the second term in (\ref{55}) vanishes and we obtain:
\begin{eqnarray}
\pi_{\rm td}|_{q_{M,\tau}}J=\left[1-{D_{\tau}q_{M,\tau}D_fS|_{q_{M,\tau}}\over
D_fS|_{q_{M,\tau}}D_{\tau}q_{M,\tau} }\right]D_Mq_{M,\tau}mJ.\label{56}
\end{eqnarray}
The equation for $M$ corresponding to (\ref{56}) has the form:
\begin{eqnarray}
\dot{M}=m\pi_{\rm td}|_{q_{M,\tau}}J(q_{M,\tau})=m\left[1-{D_{\tau}q_{M,\tau}D_fS|_{q_{M,\tau}}\over
D_fS|_{q_{M,\tau}}D_{\tau}q_{M,\tau} }\right]D_Mq_{M,\tau}mJ=mJ(q_{M,\tau}).\label{57}
\end{eqnarray}
It should be supplemented with the equation for $S$:
\begin{eqnarray}
{dS \over dt }=0,\label{58}
\end{eqnarray}
or for $\tau,$ in accordance with (\ref{54}),
\begin{eqnarray}
{d\tau\over dt}={\dot{S}-D_fS|_{q_{M,\tau}}D_Mq_{M,\tau}\dot M\over
D_fS|_{q_{M,\tau}}D_{\tau}q_{M,\tau}}=-{D_fS|_{q_{M,\tau}}D_Mq_{M,\tau}\dot M\over
D_fS|_{q_{M,\tau}}D_{\tau}q_{M,\tau}},\label{59}
\end{eqnarray}
where $\dot M$ is defined in accordance with (\ref{57}). The numerator in (\ref{59}) has a simple
meaning: it is the rate of the entropy production  by dynamic equations (\ref{57}) when $\tau$ is
constant (for frozen $\tau$). Expression (\ref{59}) can be obtained from the condition of the
constant entropy for the motion on the film in accordance with (\ref{57},\ref{59}). Equations
(\ref{57},\ref{59}) describe dynamics on the film (Fig. \ref{FigDinFilm}).

Let us further assume that condition (\ref{dq}) is satisfied:
\begin{eqnarray}
q_{M,\tau}=f^*_M+\tau\Delta _{f^*_M}+o(\tau).\nonumber
\end{eqnarray}
In expressions (\ref{54},\ref{57},\ref{59}) the denominator, $D_fS|_{q_{M,\tau}}D_{\tau}q_{M,\tau}$,
is present. For $\tau \to 0$ this expression vanishes:
\begin{eqnarray}
D_{\tau}q_{M,\tau}|_{\tau=0}&=&\Delta_{f^*_M},\nonumber\\* D_fS|_{f=f^*_M}x&=&0,
\hspace{1cm}\mbox{for}\hspace{1cm} x\in \ker m,
\end{eqnarray}
$m(\Delta_{f^*_M})=0,$ therefore $D_fS|_{q_{M,\tau}}D_{\tau}q_{M,\tau}\to 0$ for $\tau \to 0.$ For
$\tau \to 0$ indeterminate forms $0/0$ appear in expressions (\ref{54}-\ref{56},\ref{58},\ref{59}).
Let us resolve the indeterminate forms and calculate the corresponding limits.

Two indeterminate forms are present:
\begin{eqnarray}
N_1={(D_{\tau}q_{M,\tau})(D_fS|_{q_{M,\tau}})D_{M}q_{M,\tau}mJ\over
D_fS|_{q_{M,\tau}}D_{\tau}q_{M,\tau} }\label{60}
\end{eqnarray}
and right hand side of equation (\ref{59}). Let us evaluate the indeterminate form (\ref{60}) with
the L'H\^{o}pital rule. We obtain:
\begin{eqnarray}
N_1(\tau)\rightarrow_{\tau \to 0} {\Delta_{f^*_M} D_f S|_{f^*_M} \pi_{f^*_M}D_fJ(f)|_{f^*_M}\over
\langle\Delta_{f^*_M}|\Delta_{f^*_M}\rangle_{f^*_M}}
\end{eqnarray}
using identity (\ref{id}), similar to (\ref{production}), we obtain:
\begin{eqnarray}
N_1(\tau){\rightarrow}_{\tau \to
0}-{\Delta_{f^*_M}\langle\Delta_{f^*_M}|\Delta_{f^*_M}\rangle_{f^*_M}\over
\langle\Delta_{f^*_M}|\Delta_{f^*_M}\rangle_{f^*_M} }=-\Delta_{f^*_M}.\nonumber
\end{eqnarray}
In such a way, for $\tau \to 0$
\begin{eqnarray}
\pi_{\rm td}|_{q_{M,\tau}}J(q_{M,\tau})\rightarrow D_Mf^*_M
mJ(f^*_M)+\Delta_{f^*_M}=\pi_{f^*_M}J(f^*_M)+(1-\pi_{f^*_M})J(f^*_M)=J(f^*_M).\label{61}
\end{eqnarray}
Similarly, after simple calculations we obtain that:
\begin{eqnarray}
{d\tau\over dt}\rightarrow 1, \: \:\mbox{for}\, \: \: \tau \rightarrow 0.\label{62}
\end{eqnarray}

The fact that for $\tau \to 0$ the action of the thermodynamic projector on $J$ becomes trivial,
$\pi_{\rm td}J=J,$ can be obtained (without calculations) from the construction of $q_{M,\tau}$ in
vicinity of zero. We have chosen this dependence in such a way that $J(q_{M,\tau})$ becomes
transverse to the film for $\tau \to 0.$ This follows from the condition (\ref{dq}). Let us
emphasise, however, that derivation of the formulas (\ref{56}-\ref{59}) themselves was not based on
(\ref{dq}), and they are applicable to any ansatz, $q_{M,\tau},$ not necessarily with the right
behavior near the quasiequilibrium (if one needs such ansatzes for anything).

\subsection{Fixed points and ``right asymptotics" for the film
equation}\label{4.3}

What is the dynamics of the film in accordance with equation (\ref{50})? A naive expectation that
$q_{M,\tau}$ tends to the stable point of equation (\ref{50}) leads to strange consequences. Stable
point (\ref{50}) is the invariant manifold $q_{M}$. On this manifold
\begin{eqnarray}
J(q_M)=D_Mq_MmJ(q_M),\label{63}
\end{eqnarray}
i.e. the projection of the vector field, $J,$ onto $q_M$ coincides with $J.$ Were the condition
$q_{M,\tau}\to q_M$ satisfied for $\tau\to \infty,$ the dynamics would become more and more
conservative. On the limit manifold $q_M,$ the entropy should be conserved. This could lead to
unusual consequences. The first of them is limited extendability backwards ``in the entropy".

Let us consider the set of points $M_{-\tau}$ for given $M.$ Because of the existence of the limit,
$T_{\tau}M_{-\tau}\to q_M,$ for $\tau \to 0,$ the difference, $S(M)-S(M_{-\tau})=\Delta S_{\tau},$ is
bounded on the half-axis, $\tau\in[0;+\infty):\Delta S_{\tau}<\Delta S_{\infty}(M).$ this means that
it is impossible to get into the values of macroscopic variables, $M,$ from the quasiequilibrium
initial conditions, $M_1,$ for that $S(M)-S(M_1)>\Delta S_{\infty}(M).$ Assuming additionally a
smoothness of $q_M$ and $M_{-\tau},$ we see that it is impossible to get into
$\varepsilon-$neighborhood of the quasiequilibrium state, $M^*,$ (over macro-variables) from the
outside, from the quasiequilibrium initial conditions $M_0,$ if $S(M_0)<S_{\varepsilon},$ where
$S_{\varepsilon}$ is $\varepsilon$ dependent threshold of the entropy. Thus, possible stable points
of the equation (\ref{50}), regardless of their obvious interest, likely demonstrate exotic
possibilities. The following ``right asymptotics" correspond to our qualitative expectations for
large $\tau.$ Namely, it is expected that for the quite large $\tau,$ $\dot{M}$ becomes, within a
good precision, a function of $M,$ and later does not depend on $\tau:$
\begin{eqnarray}
m(J(q_{M,\tau}))\rightarrow\dot{M}(M);\label{64}
\end{eqnarray}
with the entropy production:
\begin{eqnarray}
\sigma(q_{M,\tau})=D_MS(M)mJ(q_{M,\tau})\rightarrow \sigma(M)>0,\label{65}
\end{eqnarray}
and, correspondingly, $S(q_{M,\tau})\to -\infty,\,\tau \to \infty.$

Already simple examples (linear in $J$) demonstrate that it is not so simple to construct such an
asymptotic. Moreover, for reasonably built systems it probably does not exist. Indeed, let $J(q)=Lq$,
we search for ``the right asymptotic" in the form $q_{M,\tau}=a(M)+\tau b(M)+o(1).$ We obtain:
\begin{eqnarray}
mb=mLb&=&0,\nonumber\\*ma(M)&=&M,\nonumber\\* Lb(M)-D_Mb(M)mL(a(M))&=&0,\nonumber\\*
La(M)-D_Ma(M)mL(a(M))&=&b(M).\label{66}
\end{eqnarray}

Acting with the operator $mL$ on the first equation, we obtain $mL^2b(M)=0;$ further, acting with the
operator $mL^2,$ we obtain $mL^3b(M)=0,$ and so on.

Thus,
\begin{eqnarray}
b(M)\subset \bigcap^{\infty}_{k=0}\ker mL^k=H.\label{67}
\end{eqnarray}
Space $H$ is $L-$invariant, therefore, it is possible to pass from the initial dynamics,
$\dot{f}=Lf,$ to the dynamics in the factor-space. This does not change the dynamics of macroscopic
variables because of the definition of $H$ (\ref{67}).

In such a way, instead of the right asymptotic equations, (\ref{66}) leads us again to the equation
of the invariant manifold ($b=0,$ $a(M)$ determines the invariant manifolds.)

\subsection{Coarse-graining projector}\label{4.4}

A construction of an exact projection of the microscopic dynamics on the macroscopic is meaningless,
it has meaning only as an intermediate result. Really, generically, such a projection (finite segment
of the trajectory $M(t)$) contains practically all the information about the Liouville equation. This
is a bit too much.

Moreover, there are no invariant manifolds with the dissipative dynamics for the finite-dimensional
conservative systems. The conclusion is: every time explicitly, or sometimes implicitly,
coarse-graining, or replacing of the system with something different, takes place.

For example, there is no invariant manifold for the Liouville equation parameterized with the
one-particle distribution function with dissipative dynamics on this manifold. The derivation of the
Boltzmann equation requires some limit transitions.

A few ways of coarse-graining are known, but essentially only two exist. The first one is related to
distinguishing a manifold, $M,$ and a projector, $\Pi,$ on it; the manifold, $M,$ with the projector,
$\Pi,$ separate the ``microscopic" ($\ker \Pi$) from the macroscopic, $M.$ In particular, $m(f)=m(\Pi
f).$ Then, a ``new microscopic dynamic" is parameterized: instead of $\dot{M}=m(J(f))$ the equality
\begin{eqnarray}
\dot{M}=m(J(\Pi f))\label{68}
\end{eqnarray}
is used.

Equation (\ref{68}) determines, for example, quite different system of relations in the chain of
derivatives $M,$ $\dot{M},$ $\ddot{M}$ (original system can contain no such relations). For
construction of ``right asymptotics" relation (\ref{66}) is much more appropriate then the correct
initial relation: for example, instead of $mL^k$ in (\ref{67}) one can get $m(L\Pi)^k$, and the
kernel $(L\Pi)^k$ always contains the kernel $\Pi.$

If the coarse-graining projector, $\Pi,$ and manifold, $M,$ are postulated, then the {\it hypothesis
of existence of the thermodynamic limit} consists of existence of the limit:
\begin{eqnarray}
\Pi q_{M,\tau}\rightarrow f^\#_M\hspace{1cm}\mbox{for}\,\tau\rightarrow \infty.\label{69}
\end{eqnarray}

Nevertheless, one should not expect a precise fulfillment of equality (\ref{69}), this is still the
case of exact projection (on $M$). For $\tau \to \infty$ one can expect realistically only the
smallness of the remainder:
\begin{eqnarray}
\delta (\tau,M,N)=\Pi q_{M,\tau}-f^\#_M,\label{70}
\end{eqnarray}
 and its elimination for $N\to
\infty,$ where $N$ is number of particles.

For a more precise formalization of this condition one should estimate $\delta$ under $\tau \to
\infty,$ for example, such as:
\begin{eqnarray}
\varepsilon(M,N)=\sqrt{\overline{\lim_{\tau\to\infty}}{1\over \theta}\int_0^{\theta}\|\delta
(\tau,M,N)\|^2d\tau}\label{71}
\end{eqnarray}
and, further, investigate the thermodynamic limit $N\to \infty$.

So far, however, the number of particles was not taken into consideration, and we dealt with only one
fixed system. In this case we can't help but to assume that the value of $\varepsilon(M)$ (\ref{71})
is ``sufficiently small". We notice that the problem of determining the dynamic for the thermodynamic
limit of infinite systems is a very difficult problem. For infinite system even determining the
energy, entropy, and other characteristics is not clear. We are talking not about the limit does not
exist (the number of particles is always finite), but about the asymptotics for large $N$, therefore,
strictly speaking, one needs not only to know the limits, but to estimate the reminding terms too.

The second coarse-graining method consists of a decomposition of the system into small subsystems,
and introduction of two incomparable time scales: micro and macro. The main assumption is that during
an arbitrarily small macroscopic time period a small part of the system passes the micro-evolution
within an infinitely long time. This leads to a quasi-chemical description: within each period a
number of elementary atomic processes (events) takes place. For example, the derivation of the
Boltzmann equation in frames of all formalisms is, in fact, reduced to this. We will return to
considering this approach, but for now we'll concern ourselves with the coarse-graining projector for
the film.

\subsection{Choice of the coarse-graining projector,
 and layer-by-layer linearization}\label{4.5}
The simplest choice of the coarse-graining projector is
\begin{eqnarray}
\Pi(f)=(M(f),\dot{M}(f))=(mf,mJ(f)).
\end{eqnarray}

For many problems, for example, to investigate the invariance defect (\ref{68}) it is not necessary
to place the manifold $M$ in the initial space $E$, it is sufficient to investigate $\dot{M}^{\#}_M.$

In the cases when one needs, after all, to have corresponding elements of $E$, a good choice could be
the quasiequilibrium manifold corresponding to $\Pi$. The quasiequilibrium manifolds have an evident
but important property. Let
\begin{eqnarray}
f\stackrel{m_1}{\rightarrow}M_1\stackrel{m_2}{\rightarrow}
M_2,\hspace{1cm}f\stackrel{m_1m_2}{\longrightarrow}M_2,
\end{eqnarray}
be a sequence of linear mappings, where $m_1,\, m_2$ are mappings ``on", their images are whole
corresponding spaces.

Let, furthermore, ${\mathbf M}_1\subset E$ be a quasiequilibrium manifold in $E,$ corresponding to
$m_1,$ ${\mathbf M}_2$ be a quasiequilibrium manifold of macro-variables, $M_2,$ corresponding to
$m_2,$ ${\mathbf M}_{21}$ be a quasiequilibrium manifold in $E$ corresponding to $m_2m_1$. Then
\begin{eqnarray}
m_1({\mathbf M}_{21})=m_2,\hspace{0.5cm}\mbox{or}\hspace{0.5cm} {\mathbf M}_{21}=m_1^{-1}({\mathbf
M}_{21}).\label{72}
\end{eqnarray}
For the transition to the quasiequilibrium approximation this property reads simpler:
\begin{eqnarray}
U_2U_1=U_{21},\label{73}
\end{eqnarray}
where $U_i$ the  corresponding to $m_i$ procedure of the taking of the quasiequilibrium
approximation.

For each $M_2$ both the point of the quasiequilibrium, $M_1^*(M_2)$, and the linear manifold,
$m_2^{-1}(M_2)$, containing this point are defined. For each $M_1$ the quasiequilibrium,
$f^*_1(M_1)$, and the linear manifold, $m_1^{-1}(M_1)$ containing this point are defined. As well
$f^*_2(M_2),$ and containing them $(m_2m_1)^{-1}(M_2)$ are defined.

Relations:
\begin{eqnarray}
f^*_2(M_2)&=&f^*_1(M^*_1(M_2)),\nonumber\\* m_1^{-1}(m_2^{-1}(M_2))&=&(m_2m_1)^{-1}(M_2).\label{74}
\end{eqnarray}
are fulfilled.

The quasiequilibrium manifold, $f^*_2(M_2)\subset E$ parameterized by $M_2$ lies on the
quasiequilibrium manifold, $f^*_1(M_1)\subset E$ parameterized by $M_1.$ For each $M_2$ the set
\begin{eqnarray}
\{f^*_1(M_1)|f^*_1(M_1)\subseteq (m_2m_1)^{-1}(M_2)\}
\end{eqnarray}
forms the quasiequilibrium manifold in $(m_2m_1)^{-1}(M_2)$ with the set of macroscopic variables,
$m_2^{-1}(M_2),$ and the same entropy. For the projector, $\Pi(f)=(M(f),\dot{M}(f)),$ it means that
for each $M$ in the linear manifold, on that $m(f)=M,$ the quasiequilibrium manifold corresponding to
the macroscopic variables $\dot{M}(f)=mJ(f)$ (if $J(f)$ is a linear mapping on this manifold) is
defined.

The last remark leads us to an important construction named by us {\it ``layer-by-layer
linearization"}. The filed $J(f)$ could be presented in the form:
\begin{eqnarray}
J_L(f)=J(f^*_{m(f)})+D_fJ(f)|_{f^*_{m(f)}}(f-f^*_{m(f)}).\label{75}
\end{eqnarray}

The {\it ``layer-by-layer quadratic entropy"} has special importance for the theory of non-linear
equations (\ref{75}) :
\begin{eqnarray}
S_L(f)=S(f^*_{m(f)})-(1/2)\langle f-f^*_{m(f)}|f-f^*_{m(f)}\rangle_{f^*_{m(f)}}.\label{76}
\end{eqnarray}
Let us remind that the bilinear form, $\langle|\rangle_{f^*_{m(f)}},$ is generated by the negative
second differential of the entropy at the point $f.$

The layer-by-layer linearized equations allowed us to add more moment equations and construct the
quasiequilibrium approximations using the entropy (\ref{76}). This is especially important for the
moments which are time derivatives $\dot{M},\,\ddot{M}$ and so on.

Application of the layer-by-layer linearized equations (\ref{73}) together with the layer-by-layer
quadratic entropy (\ref{76}) allowed us to construct a thermodynamically consistent theory of the
moment equations for the Boltzmann equation \cite{4,MBChLANL}.

It is convenient to supplement the quasiequilibrium, $f^*_M,$ with the quasiequilibrium for
additional macro-variables $\dot{M}$,
\begin{eqnarray}
\dot{M}(f)=m(D_fJ(f)|_{f^*_{m(f)}}f),\label{77}
\end{eqnarray}
in two stages: i) supplementing by the entropy production, ii) and later by the conserving part of
the entropy.

i) We supplement $M$ by the entropy production. In the layer-by-layer linear approximation
\begin{eqnarray}
\sigma(f)=\langle\Delta_{f^*_{m(f)}}|f-f^*_{m(f)}\rangle_{f^*_{m(f)}}\label{78}
\end{eqnarray}
(as was already determined, see (\ref{sigma})). The quasiequilibrium manifold corresponding to
$\sigma$ in the layer over $f^*_M$ has the form:
\begin{eqnarray}
f^*_{M,\sigma}=f^*_M+{\sigma(f)\Delta_{f^*_M}\over
\langle\Delta_{f^*_M}|\Delta_{f^*_M}\rangle}_{f^*_M}.\label{79}
\end{eqnarray}
Quasiequilibrium projector in the layer is:
\begin{eqnarray}
\pi_{\sigma}={|\Delta_{f^*_M}\rangle\langle\Delta_{f^*_M}| \over
\langle\Delta_{f^*_M}|\Delta_{f^*_M}\rangle}_{f^*_M}\nonumber
\end{eqnarray}

ii) We distinguish in $L=DJ(f)|_{f^*_M}$ the conservative (conserving the entropy) part over $f^*_M:$
\begin{eqnarray}
L^C_M\varphi=L_M(\varphi-\pi_{\sigma}\varphi).\nonumber
\end{eqnarray}
This corresponds to the situation when we have fixed
$\sigma(f)=\langle\Delta_{f^*_M}|\varphi\rangle,$ and we consider the motion in the layer for fixed
$\sigma.$ For this:
\begin{eqnarray}
L_M\varphi=L^C_M\varphi+{L_M\Delta_{f^*_M}\langle\Delta_{f^*_M}|\varphi\rangle \over
\langle\Delta_{f^*_M}|\Delta_{f^*_M}\rangle_{f^*_M}}=L^C_M\varphi+{\sigma(\varphi) \over
\langle\Delta_{f^*_M}|\Delta_{f^*_M}\rangle_{f^*_M}}.\label{80}
\end{eqnarray}

The quasiequilibrium manifold corresponding to $L^C_M$ in the layer over $f^*_M$ could be constructed
in the following way: we search for the kernel of $L^C_M,$ in $L$ (the set of all solutions to
equations $L^C_M\varphi=0,$ $m\varphi=0$). We define it as $K.$ The orthogonal complement,
$K^{\bot},$ to $K$ in the scalar product, $\langle|\rangle_{f^*_M},$ is the corresponding manifold.
For each point from the image, $L_M,$ on $L$, $\psi\in L^C_M(L),$ there exists unique $\varphi \in
K^{\bot}_{f^*_M}$ such that $L_M\varphi=\psi.$ We define it as $\varphi=(L^C_M)^{-1}(\psi).$ As a
result, for every $\psi\in L^C_M(L)$
\begin{eqnarray}
f^*_{M,\sigma,\psi}=f^*_M+{\sigma \Delta_{f^*_M}\over
\langle\Delta_{f^*_M}|\Delta_{f^*_M}\rangle_{f^*_M}}+(L^C_M)^{-1}\psi.\label{81}
\end{eqnarray}

The second and third terms in (\ref{81}) are reciprocally orthogonal in the scalar product
$\langle|\rangle_{f^*_M}$.

\subsection{The failure of the simplest Galerkin approximations for conservative systems}\label{4.6}

The simplest approach to the problem is connected to the Galerkin approaches: one considers a
projection of the vector field, $J(f),$ onto the manifold in question and investigates the obtained
motion equations. It is not difficult to make sure that for conservative systems such an approach is
unfruitful. If the orthogonal projection, $\langle|\rangle_{f^*_M},$ is taken, then in the linear
within the layer approximation only quasiequilibrium approximations with increased number of moments
could be obtained. For the dissipative systems, in contrast, such a way leads to quite satisfactory
results. Thus, if for the Boltzmann equation and the hydrodynamic moments the invariant manifold is
to be searched in the form $f^{\#}_M=f^*_M+a(M)\Delta_{f^*_M}$, then we obtain the Navier-Stokes
equations with the viscosity calculated within the first Sonine polynomials approximation. Using
another scalar product simply leads to unphysical results.

In order to specify appearing problems, let us give an example with a linear field, $J(f)=Af,$ and
quadratic entropy, $S(f)=(1/2)\langle f|f\rangle.$ The conservativity of $J$ means that for each $f$
\begin{eqnarray}
\langle f|Af\rangle=0\label{82}
\end{eqnarray}
is fulfilled.

The quasiequilibrium subspace corresponding to the moments $M=mf$ is the orthogonal complement, $\ker
M.$ The quasiequilibrium projector, $\pi,$ is an orthogonal projector on this subspace, and does not
depend on the point. For the defect of invariance $\Delta_{f^*_M}$ we obtain:
\begin{eqnarray}
\Delta_{f^*_M}=(A-\pi A)f^*_M.\label{83}
\end{eqnarray}

Under Galerkin approximation we write
\begin{eqnarray}
q_{M,\tau}=f^*_M+a(M,\tau)\Delta_{f^*_M}.\label{84}
\end{eqnarray}

Projector of the vector field on $\Delta_{f^*_M}$ is
\begin{eqnarray}
{|\Delta_{f^*_M}\rangle\langle \Delta _{f^*_M}|\over \langle\Delta_{f^*_M}| \Delta
_{f^*_M}\rangle}.\label{85}
\end{eqnarray}

Thus, we pass from the equation of motion of the film (\ref{50}) to the Galerkin approximation for
$a(M,\tau).$
\begin{eqnarray}
&&\dot{a}=1+ \\ && a{\langle\Delta_{f^*_M}|A \Delta _{f^*_M}\rangle\over \langle\Delta_{f^*_M}|
\Delta _{f^*_M}\rangle}-a{\langle\Delta_{f^*_M}|A\pi A \Delta _{f^*_M}\rangle\over
\langle\Delta_{f^*_M}| \Delta _{f^*_M}\rangle}-a^2{\langle\Delta_{f^*_M}|A\pi A \Delta
_{f^*_M}\rangle\over \langle\Delta_{f^*_M}| \Delta
_{f^*_M}\rangle}-(D_Ma)m{Af^*_M+aA\Delta_{f^*_M}\over \langle\Delta_{f^*_M}| \Delta
_{f^*_M}\rangle}\nonumber.\label{86}
\end{eqnarray}

One can try to find the fixed points (solving $\dot{a}=0$). This is the projected invariance
equation. Due to the properties of the operator $A,$ and the self-adjoint projector, $\pi$, we obtain
for conservative systems
\begin{eqnarray}
{\langle\Delta_{f^*_M}|A \Delta _{f^*_M}\rangle}&=&0,\label{87}\\* {\langle\Delta_{f^*_M}|A\pi A
\Delta _{f^*_M}\rangle}&=&-{\langle\pi A \Delta_{f^*_M}|(\pi A^2-(\pi A)^2) \Delta
_{f^*_M}\rangle}.\label{88}
\end{eqnarray}

On the other hand, for the dissipative systems the form (\ref{87}) is negatively definite, and it is
this form that determines the Navier-Stokes equations (in the first Sonine's polynomials
approximation) for derivation of these equations from the Boltzmann equation. For the conservative
equations this main part vanishes, and the second term in equation (\ref{86}), generally speaking, is
sign-indefinite.

The failure of the Galerkin approximations is even more obvious in the equations of motions on the
film. Here everything is very simple:
\begin{eqnarray}
\dot{a}=1+a{\langle\Delta_{f^*_M}|A \Delta _{f^*_M}\rangle\over \langle\Delta_{f^*_M}| \Delta
_{f^*_M}\rangle}.\label{89}
\end{eqnarray}

For the dissipative systems under frozen $M,$ $a$ relaxes to the stable point
\begin{eqnarray}
a=-{\langle\Delta_{f^*_M}| \Delta _{f^*_M}\rangle\over \langle\Delta_{f^*_M}| A\Delta
_{f^*_M}\rangle}.\label{90}
\end{eqnarray}

This fixed point is ``the leading order term" in the solution of the invariance equation, $\dot{a}=0$
(\ref{86}).

For the conservative systems $\dot{a}=1.$ This result was evident beforehand from the entropy
production formula (\ref{production}), and
\begin{eqnarray}
-S(f)=(1/2)\langle f|f\rangle =(1/2)\langle \pi f|\pi
f\rangle+(1/2)\langle(1-\pi)f|(1-\pi)f\rangle.\nonumber
\end{eqnarray}

\subsection{Possible ways beyond the simplest Galerkin
approximations}\label{4.7}

The first way is an application of the projection operators methods \cite{5}. The film equation
(\ref{48}) is considered for two sets of ``variables": slow ``macro-variables", $\Pi q,$ and rapid
``micro-variables", $(1-\Pi)q$ (where $\Pi$ is a coarse-graining projector, see subsection
\ref{4.4}).

Next the rapid variables are eliminated, and the equation with retardation for the slow variables is
written. This formally exact equation becomes tractable only after a sequence of additional
approximations (``short memory", ``Markovian models" etc.). The method is applicable to linear
(linear within the layer) vector fields, $J(q)=L_Mq$. The main problem is the computation of the
coefficients including averaging along the trajectories of the rapid motion.

The second way is an introduction of the dissipative part (using the thermodynamic limit) into the
vector field, $J(q_{M,\tau}).$ One adds into (\ref{50}) either a ``relaxation" operator
\begin{eqnarray}
-\gamma(q_{M,\tau}-f^*_M),\label{91}
\end{eqnarray}
or operator $\gamma P,$ simulating a random process. For example, if $f$ is a function on a space
$X,$ then the typical form of $P$ with the ``detailed equilibrium" is $f^*_M$
\begin{eqnarray}
P(f)=\int Q(x,x')\left({f(x')\over f^*_M(x')}-{f(x)\over f^*_M(x)}\right)dx'\label{92}
\end{eqnarray}
with non-negative kernel $Q(x,x')\geq 0,$ $\int Q(x,x')dx'\equiv 0.$

As a result, the system becomes dissipative, and one can construct for it invariant manifolds that
are stationary solutions for the film equation. They could be found either as a sequence \cite{6,7}
or, more effectively, based on the Newton method with the incomplete linearization \cite{8,9}. It
becomes possible to use the Galerkin approximations, and so on.

After this one makes the transition to the thermodynamic limit. It is suggested that the
thermodynamic limit exists for the found invariant manifold, $q^{\#}_M(\gamma)$, and, if later
$\gamma$ tends to zero, that a finite limit, $\Pi q^{\#}_M(\gamma)$ exists. This limit is suggested
for the definition of the macroscopic variables $M.$

In some problems of dissipative kinetics (namely, in the problem of initial layer for the Boltzmann
equation) it was found to be effective to approximate the trajectories by segments (with further
smoothing and corrections, or without them). These segments were constructed in the following way:
the initial direction of motion was taken, and $f$ evolved along this direction for as long as it was
possible to conserve the smooth evolution of the entropy. Further, the procedure was repeated from
the obtained point (for details see \cite{10,11}).

Unfortunately, in the problem of the initial layer for the conservative systems there are no stop
points during the motion along the straight line (more precisely, the start of the motion itself can
be considered as a stop point because under the linear approximation the relation (\ref{87}) is
valid). In the initial layer for the dissipative systems the motion of the system along the straight
line $x=\tau \Delta$ in any case increases the entropy. For the conservative systems one needs to
``rotate the phase", and the models of motion are to be arcs of ellipses (in linear space), or the
constant entropy lines, instead of the straight lines. In the film problem, as even the simplest
examples show, the simplest good model is a general conic section. A simple example: $J(f)=Af,$ $A$
is generator of rotation around the axis with direction $\vec{r}=\vec{e}_x+\alpha \vec{e}_y,$ $M=x,$
the film is the lateral surface of the cone, obtained by rotation of the quasiequilibrium manifold,
the axis $\{x\vec{e}_x\},$ around the axis $\{\varphi \vec{r}\}.$ For $\alpha <1$ the curve
$q_{M,\tau}$ is an ellipse, for $\alpha>1$ it is a hyperbole, for $\alpha =1$ it is a parable.

\subsection{The film: Second order Kepler models}\label{4.8}

The curve $q_{M,\tau}$ is a section of two manifolds: one of them is the result of motion of the
quasiequilibrium manifold along the vector field $J(f)$, and another one is the linear manifold
$f^*_M+\ker m.$

Already in the finite-dimensional space, and under linear approximation ($J$ is linear, $S$ is
quadratic) we have an interesting geometrical picture: quasiequilibrium manifold is an orthogonal
complement to $\ker m,$ $A$ is the rotation generator. $(\ker m)^{\bot}$ is rotated under action of
$e^{A\tau},$ unknown curve is the section:
\begin{eqnarray}
(f^*_M+\ker m)\bigcap e^{AR_{+}}(\ker m)^{\bot},\label{93}
\end{eqnarray}
where $R_{+}=[0;\infty),$ $f^*_M\in (\ker m)^{\bot}.$

\begin{figure}[p]
\begin{centering}
\includegraphics[width=150mm, height=170mm]{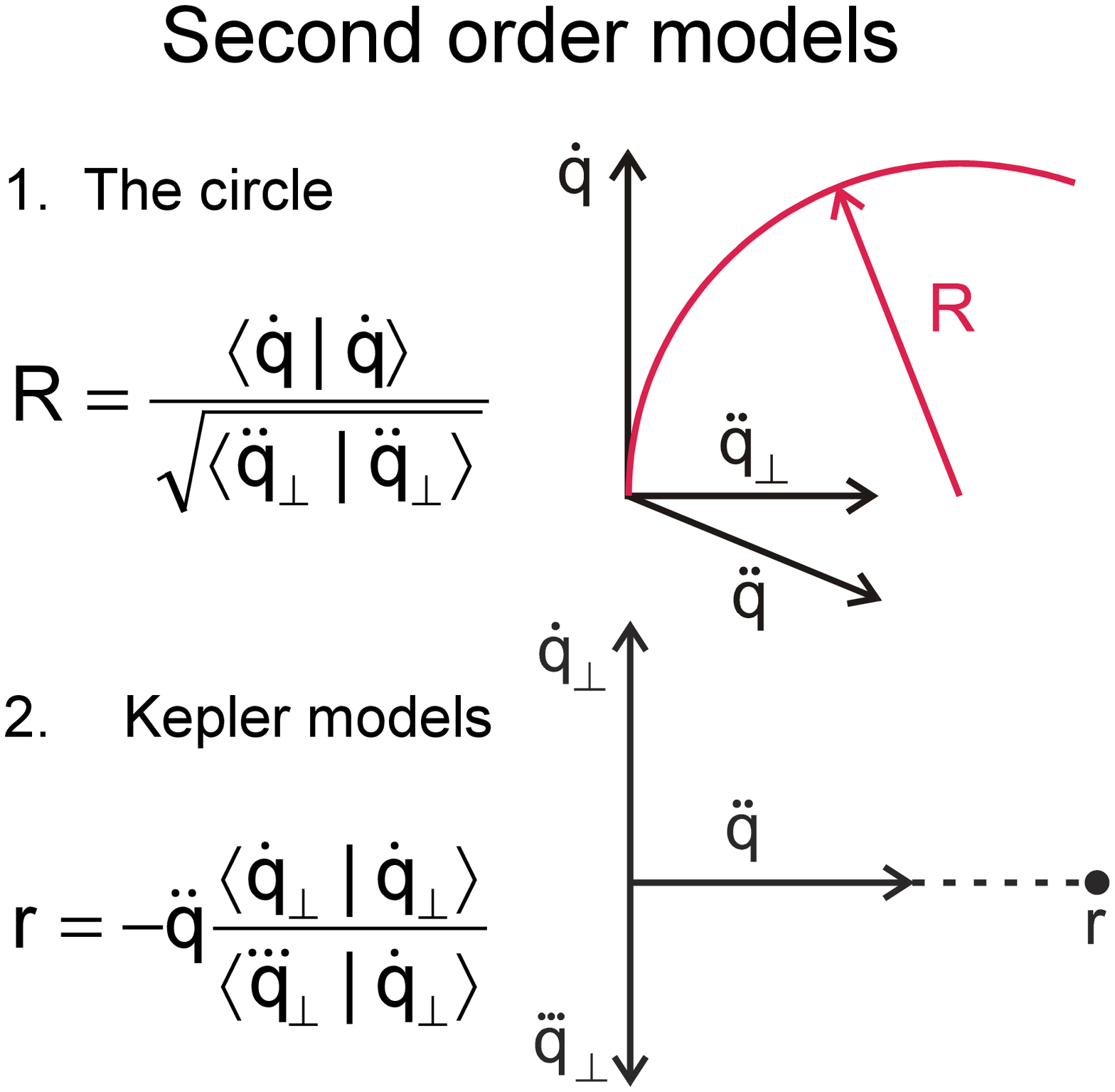}
\caption{The definition of the second-order models.} \label{FigKep}
\end{centering}
\end{figure}

The simplest model motion is a second order curve. However, it is not sufficient to know the first
and the second derivatives. We need information about the third derivative. If we consider the curve
$q_{M,\tau}$ as a trajectory in the Kepler problem, then the location, $r,$ of the center of
attraction (repulsion) is (Fig. \ref{FigKep}):
\begin{eqnarray}
r=q_0-\ddot{q}{\langle \dot{q}_{\bot}|\dot{q}_{\bot}\rangle \over \langle \stackrel{\dots}{q}
|\dot{q}_{\bot}\rangle },\label{94}
\end{eqnarray}
where $r_0$ is the initial point where all the derivatives are taken. The force is:
\begin{eqnarray}
F&=&\alpha{r-q\over \langle r-q|r-q\rangle^{3/2}};\nonumber\\*
\alpha^2&=&\langle\ddot{q}|\ddot{q}\rangle \langle r-q|r-q\rangle^2=\langle
\ddot{q}|\ddot{q}\rangle^3{\langle \ddot{q}_{\bot}|\ddot{q}_{\bot}\rangle^4\over \langle
\stackrel{\dots}{q}|\dot{q}_{\bot}\rangle^4};\label{95}
\end{eqnarray}
\begin{eqnarray}
\begin{array}{lll}
&\alpha > 0 \hspace{1cm}\mbox{(attraction)} & \mbox{ if}\hspace{0.5cm} \langle
\stackrel{\dots}{q}|\dot{q}_{\bot}\rangle <0 ;\\ &\alpha < 0 \hspace{1cm}\mbox{(repulsion)} &\mbox{
if}\hspace{0.5cm} \langle \stackrel{\dots}{q}|\dot{q}_{\bot}\rangle>0.
\end{array}
\end{eqnarray}

It is necessary to specifically note that the Kepler problem defines an approximation of the
trajectory $q_{M,\tau}$, but not the dependence on $\tau.$

An important question is the finiteness of the film. Is the modeling motion finite?

The answer is simple in terms of the Kepler problem \cite{12}:
\begin{eqnarray}
{\|\dot{q}\|^2\over 2}<{\alpha \over \|r-q_0\|},\nonumber
\end{eqnarray}
or
\begin{eqnarray}
{\|\dot{q}\|^2|\langle\dot{q}_{\bot}|\stackrel{\dots}{q}\rangle|\over 2
\|\dot{q}_{\bot}\|^2\|\ddot{q}\|^2}<1.\label{96}
\end{eqnarray}
Here $\|\,\| = (\langle|\rangle_{f^*_M})^{1/2}$ is the norm in the entropic scalar product, as it is
usual.

\subsection{Minimal second order models: entropic parable, and entropic circle}\label{4.9}

In accordance with the film equation (\ref{50}), the following derivatives
\begin{eqnarray}
\dot{q}_{M,\tau}&=&\partial q_{M,\tau}/\partial \tau;\nonumber\\* \ddot{q}_{M,\tau}&=&\partial^2
q_{M,\tau}/\partial \tau^2;\nonumber\\*\stackrel{\dots}{q}_{M,\tau}&=&\partial^3 q_{M,\tau}/\partial
\tau^3;\nonumber
\end{eqnarray}
contribute to the construction of the second order Kepler models.

There is a rougher construction leading to the two distinguished simplest second order models that
uses only two derivatives. One of them is finite (the entropic circle), another one is infinite (the
entropic parable). Both could be constructed for every point of the film (if $\ddot{q}_{\bot}\neq 0,$
otherwise, all second order models turn into straight lines). The circle was already used by us in
subsection \ref{stability} in order to estimate stabilizating value of $\sigma.$ Let us remind that:
\begin{eqnarray}
R={\langle\dot{q}|\dot{q}\rangle\over \sqrt{\langle\ddot{q}_{\bot}|\ddot{q}_{\bot}\rangle}}
\end{eqnarray}
where $\dot{q}=\partial q_{M,\tau}/\partial \tau$,  $\ddot{q}=\partial^2 q_{M,\tau}/\partial \tau^2$
\begin{eqnarray}
\ddot{q}_{\bot}=\ddot{q}-{\dot{q}\langle\dot{q}|\ddot{q}\rangle\over
\langle\dot{q}|\dot{q}\rangle}\nonumber
\end{eqnarray}
$\langle |\rangle$ is the entropic scalar product corresponding to the extension of the entropy at
the point $q_{M,\tau_0},$ or, for the linear (as well as linear within the layer) systems, at the
point $f^*_M.$

The concentric motion could be presented as:
\begin{eqnarray}
q_{M,\tau}-q_{M,\tau_0}&=&\dot{q}{R\over \|\dot{q}\|}\sin \left({\|\dot{q}\|\over
R}(\tau-\tau_0)\right)+\ddot{q}_{\bot}{R\over \|\ddot{q}_{\bot}\|}\left(1-\cos
\left({\|\dot{q}\|\over R}(\tau-\tau_0)\right)\right)\nonumber\\*
&=&\dot{q}\sqrt{{\langle\dot{q}|\dot{q}\rangle\over
\langle\ddot{q}_{\bot}|\ddot{q}_{\bot}\rangle}}\sin
\left[\sqrt{{\langle\ddot{q}_{\bot}|\ddot{q}_{\bot}\rangle \over \langle\dot{q}|\dot{q}\rangle
}}(\tau-\tau_0)\right]\nonumber\\* &+&\ddot{q}_{\bot}{\langle\dot{q}|\dot{q}\rangle\over
\langle\ddot{q}_{\bot}|\ddot{q}_{\bot}\rangle}\left(1-\cos
\left[\sqrt{{\langle\ddot{q}_{\bot}|\ddot{q}_{\bot}\rangle \over \langle\dot{q} |\dot{q} \rangle
}}(\tau-\tau_0)\right]\right)\label{97}
\end{eqnarray}
The parable could be constructed simpler:
\begin{eqnarray}
q_{M,\tau}-q_{M,\tau_0}=\dot q(\tau-\tau_0)+(1/2)\ddot{q}_{\bot}(\tau-\tau_0)^2,\label{98}
\end{eqnarray}
or even as:
\begin{eqnarray}
q_{M,\tau}-q_{M,\tau_0}=\dot q(\tau-\tau_0)+(1/2)\ddot{q}(\tau-\tau_0)^2,\label{99}
\end{eqnarray}

The difference between (\ref{98}) and (\ref{99}) is this: in the formula (\ref{98}) the angle between
$\partial q_{M,\tau}/\partial \tau|_{\varepsilon=\tau_0}$ and $\partial q_{M,\tau}/\partial \tau$
tends to $\pi/2$ for $\tau \to \infty,$ in the model (\ref{99}) this occurs too if $\langle
\dot{q}|\ddot{q} \rangle\neq 0.$

{\it Note.} In expressions (\ref{97}-\ref{99}) $\tau$  coincides with the true $\tau$ of the motion
on the film only in the zeroth and the first orders. For the further calculations it could be
necessary to recalculate $\tau$ using true values of $\dot{q}$ in the projection on the trajectory.
This is discussed below.

\subsection{The finite models: termination at the horizon points}\label{4.10}

In order to construct a step-by-step approximation it is necessary to be able to solve two problems:
the choice of the direction of the next step, and the choice of the value of this step.

\begin{figure}[p]
\begin{centering}
\includegraphics[width=150mm, height=200mm]{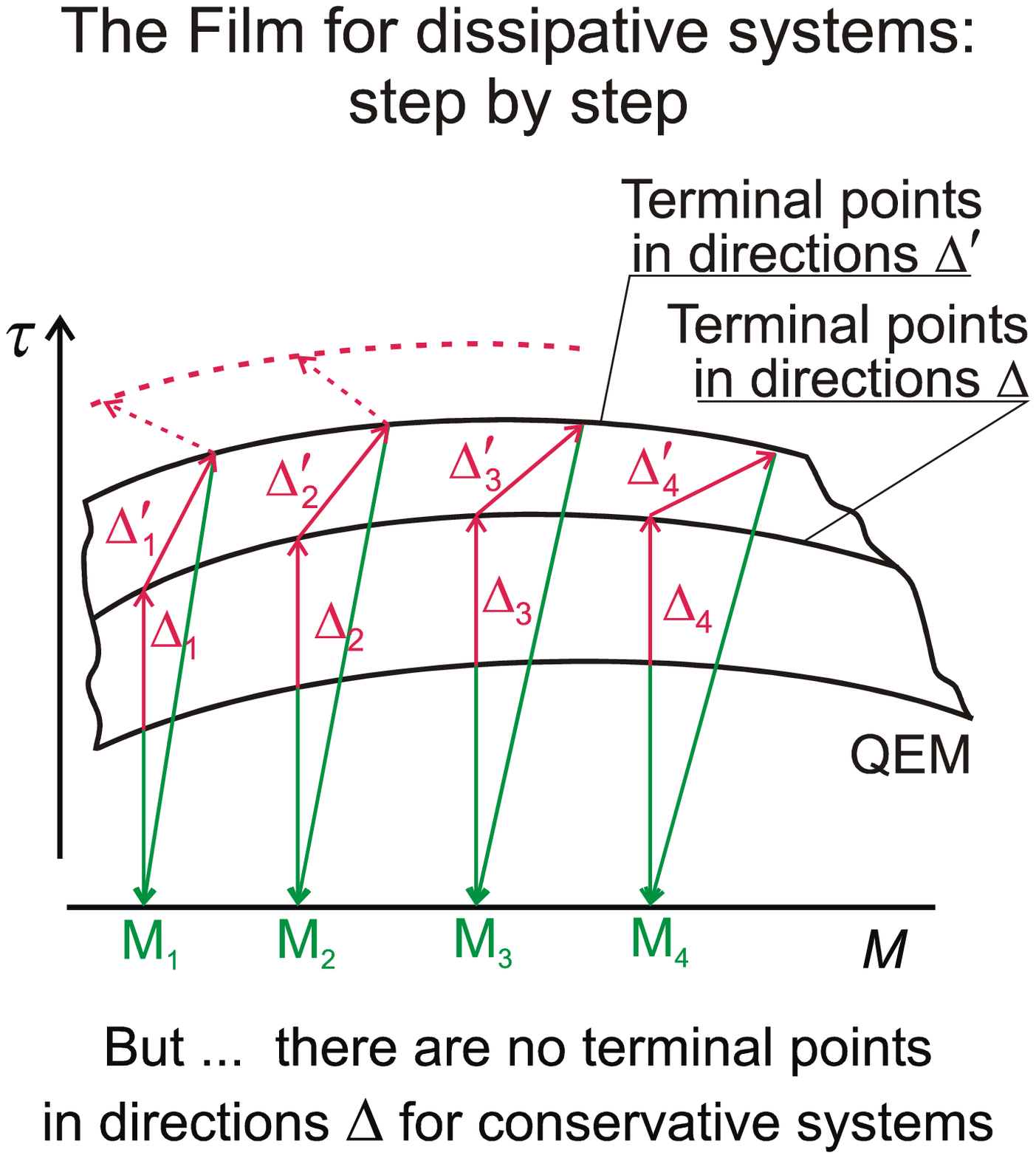}
\caption{The stepwise construction of the film for dissipative system. First-order models: The motion
along the defect of invariance.} \label{FigFDis}
\end{centering}
\end{figure}

If the motion $q_{M,\tau}$ is along the straight line (dissipative systems), the direction of the
next step is $\dot{q}_{M,\tau_0}$ (let us remind that $\dot{q}_{M,\tau_0}$ is the defect of the
invariance of the manifold $q_M=q_{M,\tau_0}$ under fixed $\tau=\tau_0$), and the value of the step
should be taken in the direction to the stable point: to the point where direction of
$\dot{q}_{M,\tau}$ becomes orthogonal to initial one, $\dot{q}_{M,\tau_0}$ (Fig. \ref{FigFDis}).
Naturally, the current direction of $\dot{q}_{M,\tau}$ is calculated with (\ref{50}), but
approximately, with the frozen projector ($D_Mq_{M,\tau_0}$ instead of $D_mq_{M,\tau}m$).

For the conservative systems we have chosen the second order models instead of the linear ones. For
finiteness of the models we need to define the moments of stop. It is suggested to operate in a
manner similar to the case of the dissipative systems: to stop at the moment when the direction of
motion is orthogonal to the initial one. In this case we will take the direction of motion along the
model.

Thus, if $q_{M,\tau_0}$ is a starting point of motion, and $\tilde{q}_{M,\tau_0+\theta}$ is a motion
on the finite second order model, then condition for the transition to the next model is
\begin{eqnarray}
\langle\dot{q}_{M,\tau_0}|{d\tilde{q}_{M,\tau_0+\theta}\over d \theta }\rangle=0\label{100}
\end{eqnarray}
(in the entropic scalar product).

Let us call {\it the horizon points} such points, $q_{M,\tau_0+\theta_0},$ where the scalar product
(\ref{100}) for the first time becomes zero (for $0\leq \theta <\theta_0$ this scalar product is
positive). This notion is motivated by the fact that for $\theta
>\theta_0$ the motion on the second order model ``disappears behind
the horizon", and its orthogonal projection on the line parallel to $\dot{q}_{M,\tau_0}$ starts to
move back passing the same points for the second time.

\begin{figure}[p]
\begin{centering}
\includegraphics[width=150mm, height=210mm]{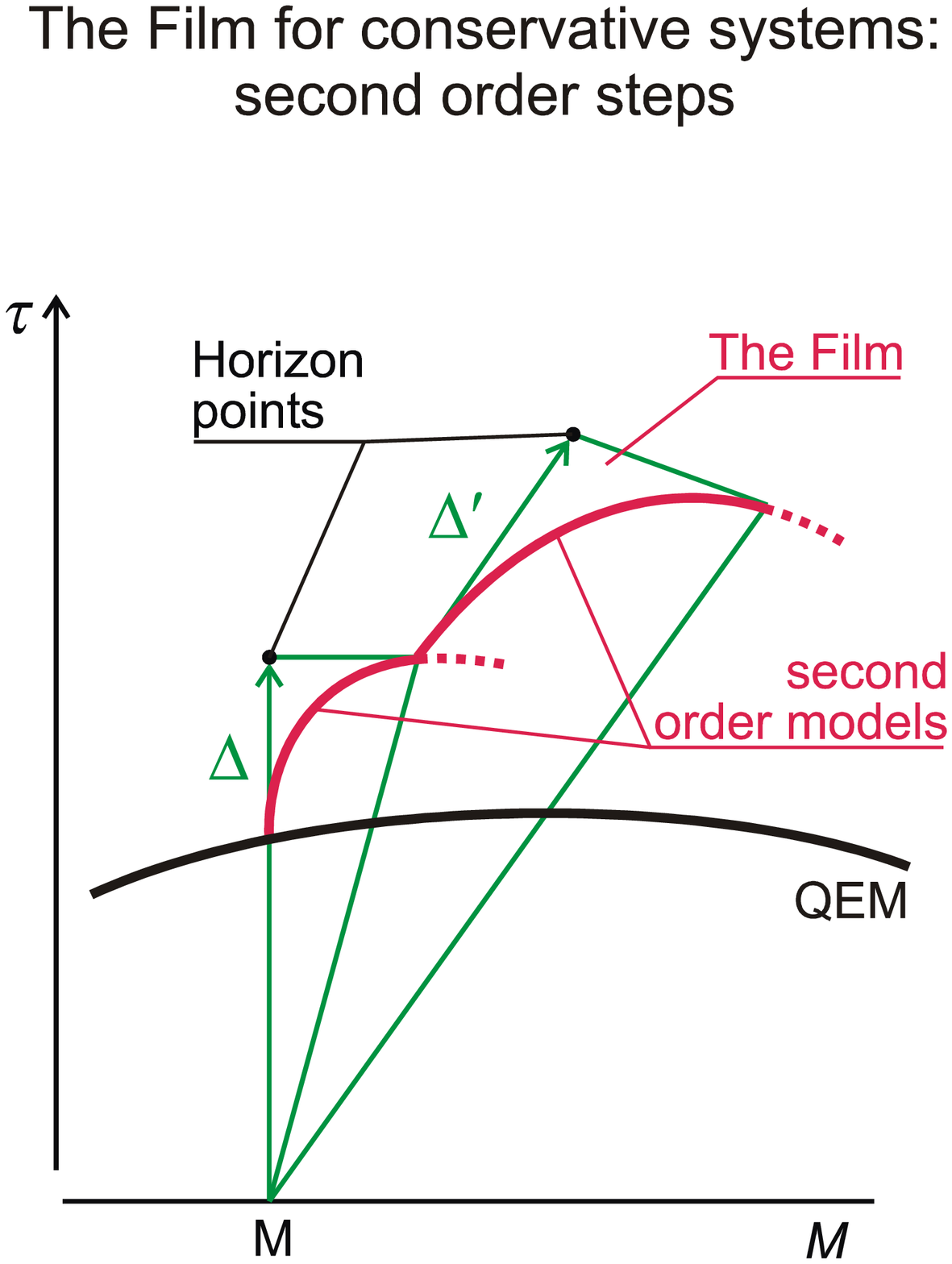}
\caption{The stepwise construction of the film for dissipative system. Finite second-order models:
The motion starts in the direction of the defect of invariance, and stops when the direction of
motion becomes orthogonal to the defect of invariance.} \label{FigFCons}
\end{centering}
\end{figure}

The convention about the change of the model in the horizon points seemed quite natural. The
following sequence of calculations suggests itself (Fig. \ref{FigFCons}):

1) we pose that $q_{M,0}=f^*_M;$

2) we calculate $\dot{q}_{M,0},$ $\ddot{q}_{M,0},$ $\dots$ in accordance with equation ({50});

3) we construct the (finite) second order models, $q_{M,\theta};$

4) we find the horizon points, $q_{M,\theta_0(M)};$

5) then we take the manifold of the horizon points as a new initial manifold, and so on.

At the first glance, this sequence contradicts the original statement of the film problem. The
manifold $q_{M,\theta_0(M)}$ does not have the form of $q_{M,\tau}$ for a fixed $\tau$ and is not a
shift of the quasiequilibrium manifold by the given time along the true microscopic equations of
motion.

The second difficulty was already mentioned: the time of motion along the modeling curve does not
coincide with the true time, $\tau.$ More precisely, it coincides only within the second order.
However, global, not local approximation are consructed. Therefore, global corrections to the time,
or ways to circumvent these corrections, are required.

The following two sections are devoted to the elimination of these difficulties.

\subsection{The transversal restart lemma}\label{4.11}

Let $q_{M,\tau}$ $(\tau\in [0;+\infty))$ be the solution to (\ref{50}) under initial condition
(\ref{51}) (the film). We call {\it the transverse section} of the film, $q_{M,\tau},$  the manifold,
$q_{M,\theta(M)},$ where $\theta(M)$ is a smooth function $0\leq \theta(M)\leq t<\infty.$

Let {\it the transverse condition} be fulfilled. Namely, for every finite patch, $M$, that does not
exclude equilibrium exists, $\varepsilon>0$ such that in this patch
\begin{eqnarray}
{\|J(q_{M,\theta(M)})-D_Mq_{M,\theta(M)}mJ(q_{M,\theta(M)})\|\over
\|J(q_{M,\theta(M)})\|}>\varepsilon\label{101}
\end{eqnarray}
in an appropriate (for example, entropic) norm. Let $\tilde{q}_{M,\tau}$ be the solution to
(\ref{50}) under the initial condition $\tilde{q}_{M,0}=q_{M,\theta(M)}.$ Then the following {\it
transverse restart lemma} is valid:
\begin{eqnarray}
q_{M,[0;+\infty)}=q_{M,[0;\theta(M)]}\bigcup \tilde{q}_{M,[0;+\infty).\label{102}}
\end{eqnarray}
here $q_{M,[a;b]}=\{q_{M,\tau}|\tau\in [a;b]\}$.

In order to prove\footnote{Let us remind that in the degree of generality being used there are no
proofs to the theorems of existence and uniqueness} this lemma, we notice that it is equivalent to
the following statement. For every $M$ the segment of the trajectory, $T_{\tau}f^*_M$
$(\tau\in[0;t]),$ crosses the manifold $q_{M,\theta (M)},$ and only once.

In order to demonstrate the unicity of the section, we consider the film in another coordinates, for
each point $q$ we set $M$ and $\tau,$: $q=T_{\tau}f^*_M.$

In these coordinates the transverse condition excludes pleats on $q_{M,\theta(M)}.$

In order to demonstrate the existence of the segment $T_{\tau}f^*_M$ $(\tau\in [0;t])$ for the cross
point, $q^*,$  with $q_{M,\theta(M)}$, we define in the neighborhood of the point $f^*_M$ on the
quasiequilibrium manifold the mapping into the neighborhood of this section point. Image of the point
$f^*_M$ is section of the trajectory $T_{\tau}f^*_M$ $(\tau\in [0;t])$ with the manifold
$q_{M,\theta(M)}$ in the neighborhood of $q^*.$ Due to the transverse condition, it performs an
isomorphism of the neighborhoods. Therefore, the set of $M$ for which the section of the trajectory
with $q_{M,\theta(M)}$ exists is open. Furthermore, it is closed, because the limit of section points
is a section point (and segment $[0;t]$ is compact). Obviously, it is not empty. Consequently, it is
the set of all possible $M.$

\subsection{The time replacement, and the invariance of the ther\-mo\-dy\-na\-mic
 projector}\label{4.12}

Let the film be constructed as $\tilde{q}_{M,\theta},$ where relation between $\theta$ and $\tau$ is
unknown; $\tau=\tau(M,\theta),$ $\theta=\theta(M,\theta),$ in order to determine this functions one
needs to solve the equation obtaining from (\ref{50}) with substitution
$q_{M,\tau}=\tilde{q}_{M,\theta(M,\tau)}$ (and projection, therefore, $\tilde{q}$ is only an
approximation). The calculation itself does not contain principal difficulties. However, a question
arises: is it possible to escape the reverse replacing of time for the derivation of the kinetic
equations? Another words, could we use the constructed geometrical object, the film, without an exact
reconstruction of the time, $\tau,$ on it?

For positive answer to this question  it is sufficient to demonstrate that the equations of motion,
constructed with the thermodynamic projector (\ref{57}-\ref{59}), describes the same motion on the
film after the time replacement.

This property of the thermodynamic projector is evident: deriving equations (\ref{57}-\ref{59}), we
did not use that $\tau$ is the ``true time" from the equation (\ref{50}), and made the local
replacement of variables, passing from $\Delta M,$ $\Delta \tau$ to $\Delta M,$ $\Delta S.$

In such a way, the thermodynamic projector is invariant with respect to the time replacement, and,
constructing equations of motion, it is not necessary to restore the ``true time".

Results of this, and previous sections allow to apply the sequence of operations suggested in
subsection \ref{4.10}.

\subsection{Correction to the infinite models}\label{4.13}

Let an infinite model $q_{M,\theta},$ $(\theta \in [0;+\infty)),$ $q_{M,0}=f^*_M$ be constructed for
the film. Actually, it means that an approximation is constructed for the whole film $q_{M,\tau}$
(but not for its initial fragment, as it was for the finite models). Naturally, there arises a
problem of correction to this initial approximation, and, in general, construction of a step-by-step
computational procedure.

The thermodynamic projector on the film is defined (\ref{56}). Correspondingly, the invariance defect
of the film is determined too
\begin{eqnarray}
\Delta q_{M,\theta}=(1-\pi_{\rm td}|_{q_{M,\theta}})J(q_{M,\theta})=
\left[1-\left(1-{D_{\theta}q_{M,\theta}D_fS|_{q_{M,\theta}} \over
D_fS|_{q_{M,\theta}}D_{\theta}q_{M,\theta}}\right)D_Mq_{M,\theta}m\right]J(q_{M,\theta})\label{103}
\end{eqnarray}

It is easy to verify, that if $q_{M,\theta}$ is a solution to (\ref{50}), then $\Delta
q_{M,\theta}\equiv 0.$

Subsequently we calculate the corrections to $q_{M,\tau}$ using an iterative method for the manifold
with edge (see Appendix).

Generally speaking, one could (and should) calculate these corrections for the finite models.
However, the infinite models are distinguished, because they require such corrections.

\subsection{The film, and the macroscopic equations}\label{4.14}

Let a film be constructed. What next? There are two routes.

i) Investigation of the conservative dynamics of $``N+1"$ variables, where $``N"$ is moments, $M,$
and $``+1"$ is $\tau$ on the film;

ii) Derivation of the macroscopic equations for $M.$

Actually, the second route is more desirable, it leads to the usual classes of equations. The first
one, however, is always available, because the film exists always (at least formally) but the
existence of equations for $M$ is not guaranteed.

The route of obtaining equations for $M$ is is the same one, suggested by us
\cite{9},\cite{17}-\cite{20} following Ehrenfest \cite{Ehrenfest}, and Zubarev \cite{22}.

i) One chooses a time $T$.

ii) For arbitrary $M_0$ one solves the problem of the motion on the film (\ref{57}), (\ref{59}) under
initial conditions $M(0)=M_0$, $\tau(0)=\tau_0$ on the segment $t\in [0;T].$ The solution is
$M(t,M_0).$

iii) For the mapping $M_0\to M(T)$ the system $dM/dt=F(M)$ is constructed. It has the property that
for its phase flow, $\theta_t(M),$ the identity
\begin{eqnarray}
\theta_T(M_0)\equiv M(T,M_0)\label{104}
\end{eqnarray}
is satisfied, for defined $T$ and all $M_0.$ This is a natural projector again (see (\ref{match}),
and whole section \label{3.1}).

In this sequence of actions there are two nontrivial problems: solution to the equations on the film,
and reconstruction of the vector field by transformation of the phase flow, $\theta_T,$ under fixed
$T.$

The natural method for solving the first problem is the averaging method. The equations of motion on
the film read
\begin{eqnarray}
\dot{M}=\varepsilon P(M,\tau);\hspace{1cm}\dot{\tau}=Q(M,\tau)\label{105}
\end{eqnarray}
where $\varepsilon$ is (formally) small parameter.

Assuming that the motion of $M$ is slow, one can write down the series of the Bogoliubov-Krylov
averaging method \cite{13}. The first term of this series is a simple averaging over the period $T:$
$\tau_1(T,M)$ is solution to the equation $\dot{\tau}=Q(M,\tau)$ under fixed $M,$
\begin{eqnarray}
M_1(t,M_0)=M_0+\varepsilon t\left({1\over T}\int_0^T
P(M_0,\tau_1(\theta,M_0))d\theta\right)\label{106}
\end{eqnarray}
for $t\in [0;T]$, and
\begin{eqnarray}
M_1(T,M_0)=M_0+\varepsilon \int_0^T P(M_0,\tau(\theta,M_0))d\theta, \label{107}
\end{eqnarray}
correspondingly.

The first correction to reconstruction of the vector field, $F(M),$ by the transformation of the
phase flow, $\theta_T(M),$ is very simple too:
\begin{eqnarray}
F_1(M)={1\over T}(\theta_T(M)-M).\label{108}
\end{eqnarray}

Hence, we obtain the first correction to the macroscopic equations:
\begin{eqnarray}
\dot{M}=F_1(M)={1\over T}\int_0^Tm(J(q_{M,\tau(t,M)}))dt,\label{109}
\end{eqnarray}
where $\tau(t,M)$ is a solution to the equation (\ref{59}) under fixed $M$ (actually,
$mJ(q_{M,\tau})$ should be substituted into (\ref{59}) instead of $\dot{M}$).

The second and higher approximations are much more cumbersome, but their construction is not a
significant problem.

Let us demonstrate an explicit expression for $\dot{M}$ (\ref{109}) for the modeling motion on the
entropic circle (\ref{radius}), (\ref{97}) for the linear in layer system (\ref{75}).

The original system is
\begin{eqnarray}
\dot{f}=J(f^*_{m(f)})+L_{m(f)}(f-f^*_{m(f)}),\nonumber
\end{eqnarray}
where $L_M=(D_fJ(f))|_{f^*_M}.$

The macroscopic equations are (see also equations (\ref{firsteq})):
\begin{eqnarray}
\dot{M}&=&m(J(f^*_M))+(1/\omega)m(L_M(\dot{q})){2\omega\over \pi}\int_0^{\pi/2\omega}\sin (\omega
t)dt\nonumber\\*&+&(1/\omega^2)m(L_M(\ddot{q}_{\bot})){2\omega \over \pi}\int_0^{\pi/2\omega}(1-\cos
(\omega t))dt\nonumber\\*&=&m(J(f^*_M))+{2\omega\over
\pi}m(L_M(\dot{q}))+(1/\omega^2)(1-2/\pi)m(L_M(\ddot{q}_{\bot})),\label{110}
\end{eqnarray}
where $\dot{q}=\Delta_{f^*_M}=J(f^*_M)-\pi_{f^*_M}J(f^*_M),$ and $\pi_{f^*_M}$ is the
quasiequilibrium projector (\ref{qeproj}),
\begin{eqnarray}
\ddot{q}&=&(1-\pi_{F^*_M})L_M((1-2\pi_{f^*_M})J(f^*_M))+D_M\pi_{f^*_M}m(J(f^*_M)),\nonumber\\*
\ddot{q}_{\bot}&=&\ddot{q}-{\langle \ddot{q}|\dot{q}\rangle_{f^*_M}\over \langle
\dot{q}|\dot{q}\rangle_{f^*_M} }\dot{q}.\nonumber
\end{eqnarray}
$\langle|\rangle_{f^*_M}$ is the entropic scalar product related with quadratic approximation to the
entropy $f^*_M:$
\begin{eqnarray}
S(f)&=&S(f^*_M)+DS|_{f^*_M}(f-f^*_M)-(1/2)\langle
f-f^*_M|f-f^*_M\rangle_{f^*_M}+o(\|f-f^*_M\|^2),\nonumber\\* \omega&=&{\langle
\ddot{q}_{\bot}|\ddot{q}_{\bot}\rangle_{f^*_M}\over \langle \dot{q}|\dot{q}\rangle_{f^*_M}}.\nonumber
\end{eqnarray}
{\it Note.} In (\ref{110}), in accordance with (\ref{59}), the model of motion on the circle
(\ref{97}) is taken without recalculating the time. Such a recalculation changes the values of the
coefficients without a change in the structure of the equation: instead of $2/\pi$ and $(1-2/\pi)$,
other numbers appear.

The entropy production for equations (\ref{110}) has the form $const{\mbox{defect of invariance}
\over \mbox{curvature}}$ (\ref{firstpro}).

In general, equations such as (\ref{110}) are determined accurately to the values of the coefficients
simply by the sequence of the horizon points of the second order finite Kepler models, and
corresponding $\dot{q_i},$ ${\ddot{q}_i}$
\begin{eqnarray}
\dot{M}=m(J(f^*_M))+\sum_i(\alpha_im(L_M(\dot{q}_i))+\beta_im(L_M(\ddot{q}_i))),\label{111}
\end{eqnarray}
with $\alpha_i, \: \beta_i>0.$

The last comment on the positivity of the ``kinetic constants" $\alpha_i$ and $\beta_i$ is important,
and cannot be easily verified every time. However, in the case under consideration it follows from
the next theorem.

{\bf The theorem about the positivity of kinetic constants.} The motion on the Kepler ellipse from
start to the horizon point always satisfies the property
\begin{eqnarray}
q-q_0=\alpha\dot{q}+\beta\ddot{q};\hspace{1cm}\alpha,\beta>0,
\end{eqnarray}
where $q_0$ is a starting point, $\dot{q},$ and $\ddot{q}$ are velocity, and acceleration
correspondingly.

This theorem follows from elementary theorems about analytical geometry of second-order curves.

For the modeling motion on the circle, strictly speaking, this is not so every time. Positivity of
the coefficients is guaranteed only for $m(L(\dot{q})),$ and $m(L(\ddot{q}_{\bot})).$

Two phenomena can be related to the increase of the number of terms in (\ref{111}): i) alteration of
the kinetic constants (terms are not orthogonal to each other, therefore, new terms contribute to the
previous processes), ii) appearance of new processes.

Motion on an infinite film can lead to the stabilization of kinetic constants as functions of $M$,
but it can also lead to their permanent transformation. In the second case one has to introduce into
macroscopic equations an additional variable, the coordinate, $\tau,$ on the film.

From the applications point of view, another form of equations of motion on the film could be more
natural. In these equations kinetic coefficients are used as dynamic variables. Essentially, this is
just another representation of equations (\ref{57}), (\ref{59}). For every kinetic coefficient, $k,$
expression $dk/dt=\psi_k(\tau,M)=\varphi_k(k,M)$ is calculated in accordance with (\ref{57}),
(\ref{59}). Substitution of variables $(\tau,M)\to (k,M)$ in this equation is possible (at least
locally) if value $k$ does not stabilize during the motion on the film. Finally, we have the system
in the form:
\begin{eqnarray}
\dot{M}=m(J(f^*_M))+\sum_jk_jF_j(M);\hspace{1cm}\dot{k}_j=\varphi_j(k_j,M).\label{112}
\end{eqnarray}
For the motion starting from the quasiequilibrium state the initial conditions are $k_j=0.$

\subsection{New in the separation of the relaxation times}\label{4.15}

The classical Bogoliubov's concept about separation of the relaxation times does not agree well with
the thesis of the quasiequilibrium initial conditions.

Originally, there are no dissipative possesses in the quasiequilibrium state (the theorem of
preservation of the type of dynamics for the quasiequilibrium approximation).

The first thing that occurs during the motion out of the quasiequilibrium initial conditions is
appearance of the dissipation. It can be described (in the first non-vanishing approximation) by
equation (\ref{macro2}). It is of special importance here that there is still no separation into
processes with various kinetic coefficients. This occurs at further relaxation stages: Various
processes appear, their kinetic coefficients are determined (see, for example, (\ref{111})) (or, in
some cases, the dynamics of the kinetic coefficients is determined). And just after this the
``hydrodynamic" relaxation occurs, which is the motion of the macroscopic variables to their
equilibrium values.

Generalizing, we can distinguish three stages:

i) appearance of dissipation;

ii) branching of dissipation: appearance of various processes;

iii) macroscopic relaxation.

It is important to notice in this schema that the determination of the kinetic coefficients can occur
at both stages: at the second stage when macroscopic (hydrodynamic) relaxation can be described in
the usual form with kinetic coefficient as functions of the macroscopic parameters, as well as in the
third phase (motion on the film), when the hydrodynamic description includes dynamics of the kinetic
coefficients also.

\section{Conclusion}

To solve the problem of irreversibility we have introduced the notion of the macroscopically
definable ensembles. They are result of evolution of ensembles from the quasiequilibrium initial
conditions under macroscopic control. The quasiequilibria (ensembles of conditional maximum of the
entropy under fixed macro-variables) are intensively used in statistical mechanics after Jaynes
\cite{23}. Papers of Rosonoer and Kogan \cite{24}-\cite{26} significantly affected our initial
investigation. The primitive macroscopically definable ensembles appear as results (for $t>0$) of
motions which start from the quasiequilibrium state (at $t=0$). The hypothesis of the primitive
macroscopically definable ensembles is very important from constructive point of view: Any
macroscopically definable ensemble can be approximated by primitive macroscopically definable
ensembles with appropriate accuracy. In accordance to this hypothesis it is possible to study the one
curve for every value of macroscopic variables. These curves form the film of nonequilibrium states.

The hypothesis about the primitive macroscopically definable ensembles is real hypothesis, it can be
true or false. There is the significant difference between this hypothesis and the {\it thesis} about
macroscopically definable ensembles. The thesis can be accepted, or not, but nobody can prove the
definition, even the definition of the macroscopically definable ensembles.

Technically, the solution to the problem of irreversibility looks as follows: we can operate only
with the macroscopically definable ensembles; the class of these ensembles is not invariant with
respect to the time inversion. The notion of the macroscopically definable ensembles moves the
problem of irreversibility into a new setting. It could be called a control theory point of view. The
key question is: Which parameters can we control? It is those parameters that are fixed until ``all
the rest" come to an equilibrium. The quasiequilibrium states are obtained in such a way.

The further development of this direction must lead to an investigation of the macro-dynamics under
controlled macro-parameters. This will be a supplement of the postulated quasiequilibrium initial
conditions with an investigation of a general case of an evolution of the controlled ensembles: The
initial condition is quasiequilibrium, after which one carries out on the system by available control
influences.

The method of the natural projector allows us to construct an approximate dynamics of
macro-variables. Under tendency of the time of projection, $\tau,$ to infinity, these equations
should tend to the actual equations of macro-dynamics, if the latter exist. This a hypothesis about
their existence for the thermodynamic limit (first, the number of particles $N\to\infty,$ and then,
the time of projection $\tau\to \infty$) is the basis of the Zubarev statistical operator \cite{22}.
Here, however, we need to make a note. Frequently, physicists use objects whose existence and unicity
are not proven: solution to the hydro- gaso-dynamics, kinetic equations etc. Often, the failure to
prove the theorems of existence and unicity is treated as an absence of an adequate mathematical
statement of the problem (definition of spaces etc.). For all this, it is assumed, that substantial
obstacles either are absent, or can be distinguished separately, independently on the theorem proof
in physically trivial situations. Existence (or non-existence) of the macroscopic dynamics is a
problem of an absolutely different kind. This question is substantial: the cases of non-existence can
be found as frequently as the usual existence.

The notion of the invariant film of non-equilibrium states, and the method of its approximate
construction allow us to solve the problem of macro-kinetic even in cases when there are no
autonomous equations of macro-kinetic. The existence of the film appears to be one of the physically
trivial problems of existence and uniqueness  of solutions.

Using the Taylor's expansions of the natural projector, the first applications already have been
constructed \cite{17,18,19}. Particularly, the post-Navier-Stokes hydrodynamic, replacing the Burnett
equations, have been found. It is free from unphysical singularities \cite{17,Lar}.

The formula for entropy production $$\sigma = const{\mbox{defect of invariance} \over
\mbox{curvature}}$$ makes the geometrical sense of dissipation clear.

Nevertheless, at least one important problem remains unsolved. This is a {\it problem of undivisible
events}: For macroscopically small time small microscopic subsystems can go through ``the whole
life", from the beginning to the limit state (or, more accurate, to the limit behaviour which may be
not only a state, but a type of motion, etc.). The microscopic  evolution of the system in a small
interval of macroscopic time can not be written in the form $$\Delta f = \dot{f}\Delta t,$$ if it is
really the system with microscopic structure, and consists of a large number of microscopic
subsystems. The evolution of microscopic subsystems in a macroscopically small time $\Delta t$ should
be described as a ``ensemble of undivisible events". An excellent hint gives us the Boltzmann
equation with undivisible collisions, another good hint gives the chemical kinetics with undivisible
events of elementary reactions. The useful formalism for description such ensembles of undivisible
events is developed. It is the ``quasi-chemical" representation (see elsewhere, for example, the
review \cite{CMIM}). But the way from a general systems to such ensembles remains unclear. It is a
challenge to the following works.

\section{Appendix}
\subsection{The method of invariant manifolds}

The aim of this appendix is to give a short presentation of the method of invariant manifold,
including positive-invariant manifolds with the fixed edge.

\subsection{Construction of the invariant sections}

Let $E$ be a vector space, in the patch $U\subset E$ the vector field (microscopic system)
\begin{eqnarray}
\dot{f}=J(f),\hspace{1cm}(f\in U).\label{p1}
\end{eqnarray}
is defined.

$J$ is assumed to be smoothly continued to the closure of $U$ positively invariantly with respect to
(\ref{p1}). It means that every solution to (\ref{p1}), $f(t)$, starting under $t=0$ in $U$, is to be
found within $U$ for every $t\in [0;+\infty).$

Let $B$ be a vector field (of macroscopic variables), and a surjective mapping $m:E\to B$ is defined.

It is required to construct such a mapping
\begin{eqnarray}
M\mapsto f^{\#}_M \: \: (M\in m(U),\, f^{\#}_M\in U),\label{p2}
\end{eqnarray}
that $m(f^{\#}_M)\equiv M,$ and $f^{\#}_M$ is a positive invariant manifold of the system (\ref{p1})
(since $U$ is positive-invariant, it is sufficient to verify a local condition: the field
$J(f^{\#}_M)$ is tangent to the manifold $f^{\#}_M$ for each $M\in m(U)$).

Actually, we continue to keep such a level of strictness (unstrictness) of reasoning when such
details as topology in $E$ and $B$ etc. are not discussed. If necessary, it could be made for
particular realizations.

We call the mapping (\ref{p2}) a section, and the problem of construction of positive invariant
manifold, $f^{\#}_M,$ an invariant section problem.

It could be solved with many methods. Here are some of them:

i) the Taylor expansion over the degrees of an appropriate parameter in a neighborhood of the initial
approximation (for example the Chapman-Enskog method \cite{14});

ii) the Newton method (as in the KAM-theory \cite{15}, \cite{16} but with a incomplete linearization,
as in the original formulation of the method of invariant manifolds for dissipative systems
\cite{8}).

iii) the implementation of Galerkin approximations for each iteration.

Omitting the well known expansions of the perturbation theory, we consider the direct methods.

In order to make one step of the Newton method with incomplete linearization, we need:

i) an approximate manifold, $f^{\#}_M,$ which we call $\Omega;$

ii) projector $P$ mapping a neighborhood of $\Omega$ on $\Omega.$

For each $f^{\#}_M$ the projector, $\pi_{f^{\#}_M},$ mapping $E$ on the tangent space,
$T_{f^{\#}_M}\Omega,$ is $\pi_{f^{\#}_M}=D_fP_{\Omega}|_{f^{\#}_M}$  ($\pi$ is differential of
$P_{\Omega}$).

Usually, the projector $P_{\Omega}$ is defined in such a way that the layers (prototypes of points
$f^{\#}_M$ under projection) could be patches on affine subspaces of $E.$

For each $f^{\#}_M$ we define the invariance defect
\begin{eqnarray}
\Delta_{f^{\#}_M}=J(f^{\#}_M)-\pi_{f^{\#}_M}J(f^{\#}_M).\label{p3}
\end{eqnarray}

The invariance equation
\begin{eqnarray}
\Delta_{f^{\#}_M}=0\label{p4}
\end{eqnarray}
is solved with the Newton method with incomplete linearization: for every $M$ we search for
$\delta_{f^{\#}_M},$ such as:
\begin{eqnarray}
\left\{\begin{array}{l} P_{\Omega}(f^{\#}_M+\delta f^{\#}_M)=f^{\#}_M,\\*(1-\pi_{f^{\#}_M})
DJ(f)|_{f^{\#}_M}\delta f^{\#}_M=-\Delta_{f^{\#}_M.}
\end{array}\right.\label{p5}
\end{eqnarray}

If the layers of $P_{\Omega}$ are patches on the affine manifolds, then (\ref{p5}) is a system of
linear equations. Another form of this system is
\begin{eqnarray}
\left\{\begin{array}{l} \pi_{f^{\#}_M}\delta f^{\#}_M=0,
\\*(1-\pi_{f^{\#}_M}) DJ(f)|_{f^{\#}_M}\delta
f^{\#}_M=-\Delta_{f^{\#}_M.}
\end{array}\right.\label{p5i}
\end{eqnarray}

It should be stressed that in equation (\ref{p5}) incomplete (in contrast to the Newton method)
linearization has been used. We did not differentiate $\pi_{f^{\#}_M}$ in $\Delta_{f^{\#}_M}$ in
(\ref{p3}).

The discussion is given in (\ref{Lagrange}). We note only that for the simplest stable self-adjoin
linear systems with incomplete linearization equations (\ref{p5}) lead generically to an invariant
subspace with the largest (i.e. closest to zero) eigenvalues. In contrast, procedures with whole
lonearization lead in this case to the subspace closest to the initial approximation.

As soon as $\delta f^{\#}_M$ is found from equations (\ref{p5}), we substitute $f^{\#}_M$ for
$f^{\#}_M+\delta f^{\#}_M;$ construct new projectors and repeat the procedure.

Solution to the invariance equation (\ref{p4}) by the Newton method with incomplete linearisation can
turn into a difficult problem. In spite of their linearity, equations (\ref{p5}) cannot be solved
easily every time. One can try to simplify the problem passing from the invariance equations to the
Galerkin approximations. The simplest example is one-dimensional approximations when $\delta
f^{\#}_M=\delta_M \Delta_{f^{\#}_M},$ and equation is solved in the projection on $\Delta:$
\begin{eqnarray}
\langle\Delta_{f^{\#}_M}|(1-\pi_{f^{\#}_M})J(f^{\#}_M+\delta_M
\Delta_{f^{\#}_M})\rangle_{f^{\#}_M}=0.\label{p6}
\end{eqnarray}
The entropic scalar product $\langle|\rangle_{f^{\#}_M}=-D^2S(f)|_{f^{\#}_M}$ is used.

Solving (\ref{p6}) with the Newton method, we obtain at the first iteration:
\begin{eqnarray}
\delta M=-{\langle\Delta_{f^{\#}_M}|\Delta_{f^{\#}_M}\rangle_{f^{\#}_M}\over
\langle\Delta_{f^{\#}_M}|DJ(f)_{f^{\#}_M}\Delta_{f^{\#}_M}\rangle_{f^{\#}_M}}.\label{p7}
\end{eqnarray}

More often for the dissipative systems the denominator is negative, and this allows us to move on.
For the conservative systems the one-dimensional Galerkin approximations lead to an unsatisfactory
result, at least, in the combination with the Newton method (with the incomplete linearization).

\subsection{The entropic thermodynamic projectors}

The simplest choice of $P_{\Omega}$ is obvious:
\begin{eqnarray}
P_{\Omega}(f)=f^{\#}_M;\label{p8}
\end{eqnarray}
for each value of $f$ the values of macroscopic variables, $m(f),$ can be calculated. Based on these
values, the corresponding point with the same value of $m(f)$ on the manifold can be obtained.

However, projector (\ref{p8}) does not satisfy the physical constrains every time.

On the set $U$ a concave function, the entropy, $S$, is defined. Two kinds of systems are under
consideration: i)(\ref{p1}) dissipative, for which $dS/dt\leq 0$, in accordance with the system, ii)
conservative, for which $dS/dt=0.$ The quasiequilibrium manifolds, $f^*_M,$ are physically
distinguished. They are the solution to the variational problem
\begin{eqnarray}
S(f)\rightarrow \mbox{max},\nonumber\\*
 m(f)=M.\label{p9}\end{eqnarray}
Application of the simplest projector (\ref{p8}) leads to the fact that the vector field,
$\pi_{f^*_M}J(f^*_M),$ preserves the type of dynamics of the system on the quasiequilibrium manifold.
For the conservative $J$ it is conservative, and for the dissipative systems it is dissipative too
(with the same entropy). Such a preservation of the type of dynamics by the projector is guaranteed
only for the quasiequilibrium manifolds.

However, practically for every manifold $\Omega=\{f^{\#}_M\}$ it is possible to construct such a
projector, $P_{\Omega},$ that every $f^{\#}_M$ is a solution to the problem
\begin{eqnarray}
S(f)\rightarrow \mbox{max},\nonumber\\* P_{\Omega}(f)=f^{\#}_M\label{p10}
\end{eqnarray}
in a neighborhood of $\Omega.$ For this it would be sufficient that for every $M$ the functional
$DS(f)|_{f^{\#}_M}$ eliminate $\ker \pi_{f^{\#}_M}:$
\begin{eqnarray}
DS(f)|_{f^{\#}_M}(\ker \pi_{f^{\#}_M})=0.\label{p11}
\end{eqnarray}
Thus, in order to control the physical sense of obtained approximations, one needs to consider the
projector depending on the manifold \cite{8}.

\subsection{Method of invariant manifold for the
 positively invariant manifolds with fixed edge}

In the problem of construction of invariant sections the position of points $f^{\#}_M$ was not fixed.
Only fulfillment of condition $m(f^{\#}_M)\equiv M$ was important. There is another kind of problems
where one needs to find a positive invariant manifold with a fixed edge. Practically, this is a
problem of construction of a trajectory of the edge in accordance with (\ref{p1}) for $t\in
[0;+\infty).$

These problems include the problem of initial layer \cite{10}, \cite{11}, as well as the problem of
construction of the film of non-equilibrium states discussed in this paper.

The iterative methods described above cannot be implemented here, because they destroy the boundary
conditions on the edge of the manifold. If the invariance conditions are fulfilled on the edge of the
initial approximation, $f^{\#}_M$, accurate to the $k-$th derivative in time, then the Newton method
leads to the fact that after the $(k+1)-$th iteration the edges of the manifold will be changed (the
film tears off the edge).

In the previous paper devoted to the problem of initial layer \cite{10} we have technically overcame
this difficulty. To do this, we simulated the trajectory as an elastic beam with a rigidly fixed end.
In the mechanical equilibrium this beam had the form of an approximate trajectory. Later, it was
elastically attracted to the result of the Newton iterations. Even though that this technique allows
to avoid the separation of contact between the edge and the film, its technical artificiality forces
us to continue to search for new methods.

The application of the Picard iterative procedure allows to conserve initial conditions. For the film
equation we write: let $q^0_{M,\tau}$ be an approximation for the film, then the Picard iteration
gives:
\begin{eqnarray}
q^1_{M,\tau}=q_{M,0}+\int_0^{\tau}\left.{\partial q_{M,\tau}\over
\partial \theta}\right|_{q^0_{M,\theta}}d\theta.\label{p12}
\end{eqnarray}
Here $\partial q_{M,\Theta}/\partial \theta_{q^0_{M,\theta}}$ is the right hand side of (\ref{50})
taken at the point $q^0_{M,\theta}.$

Let us define, as usual, $\Delta_{M,\theta}=\partial q_{M,\theta}/\partial \theta
|_{q^0_{M,\theta}}-\partial q^0_{M,\tau}/\partial \theta$ as the difference of the vector field and
its projection on the approximate manifold, then the Picard iteration obtains the form:
\begin{eqnarray}
q^{1}_{M,\tau}=q^0_{M,\tau}+\int_0^{\tau}\Delta_{M,\theta}d\theta.\label{p13}
\end{eqnarray}

The Picard iteration gives a good result for small $\tau$ but can be too radical for large one. It is
possible to use the Picard iterations together with the weight which ensures essential dependence of
the correction, $q^1_{M,\tau},$ not from all $\Delta_{M,\theta}$, $0\leq \theta \leq \tau,$ but from
those within a segment $\theta \in [\tau-h;\tau].$ For example,
\begin{eqnarray}
q^{1}_{M,\tau}=q^0_{M,\tau}+\int_0^{\tau}e^{(\theta-\tau)/h}\Delta_{M,\theta}d\theta.\label{p13A}
\end{eqnarray}

Another choice of the weight function is possible. For large $\tau$ and sufficiently small $h$ and
$\Delta\neq 0$, the formula (\ref{p13A}) gives:
\begin{eqnarray}
q^{1}_{M,\tau}=q^0_{M,\tau}+h\Delta_{M,\tau}+o(h).\label{p14}
\end{eqnarray}

For small $\tau$ we obtain:
\begin{eqnarray}
q^{1}_{M,\tau}=q^0_{M,\tau}+{\tau\over k+1 }\Delta_{M,\tau}+o(\tau \Delta ).\label{p15}
\end{eqnarray}
where $k$ is order of zero of $\Delta_{M,\tau}$ at the point $\tau=0.$

Joining (\ref{p14}) and (\ref{p15}), we obtain:
\begin{eqnarray}
q^{1}_{M,\tau}\approx q^0_{M,\tau}+{h\tau\over (k+1)h+\tau }\Delta_{M,\tau}.\label{p16}
\end{eqnarray}

In all cases the question of how to choose the step arises. The simplest solution exists for
(\ref{p16}): it is possible to take the step, $h$, depending on the point:
\begin{eqnarray}
q^{1}_{M,\tau}\approx q^0_{M,\tau}+\lambda_{M,\tau}\Delta_{M,\tau}.\label{p17}
\end{eqnarray}
where $\lambda_{M,\tau}=\min \{\tau/(k+1), \: \delta_{M,\tau}\}$, and $\delta_{M,\tau}$ is to be
found from the condition of stopping in the direction of $\Delta$ (\ref{p6}).

Various combinations of the Picard and Newton iterations can generate a separate subject for
investigation. Their simplest hybridisation consists of the following. Let for each $(M,\tau)$ the
step, $\delta_{M,\tau},$ is found according to the Newton method (\ref{p5}).

Define
\begin{eqnarray}
q^{1}_{M,\tau}= q^0_{M,\tau}+{1/h}\int_0^{\tau}e^{(\theta-\tau)/h}\delta
q_{M,\theta}d\theta.\label{p18}
\end{eqnarray}

In order to have the step value on the direction $\delta q_{M,\theta}$ close to $1$ for large $\tau,$
the multiplier $1/h$ has been used. An analog to (\ref{p16}) is
\begin{eqnarray}
q^{1}_{M,\tau}= q^0_{M,\tau}+{\tau\over (k+1)h+\tau }\delta q _{M,\theta}.\label{p19}
\end{eqnarray}

The typical time scale, which separates in (\ref{p8}) and (\ref{p9}) the Picard (small times, $\tau
\ll h$), and Newton (large time, $\tau \gg h$) iterations, can be estimated by the curvature radius
(\ref{tauR}) (in general, from the relations $\|\dot{q}\|/\|\ddot{q}\|$ or
$\|\dot{q}\|/\|\ddot{q}_{\bot}\|$ ).

Further development of the methods will be determined by particular applications.

{\bf Acknowledgments.} A.N.G. is thankful to G.\ Sh.\ Fridman for stimulating discussions and
encouragement.  A.N.G. thanks Institut des Hautes \'Etudes Scientifiques, where parts of this work
have been developed, for hospitality.
 I.V.K. is thankful to the organizers of the First Mexican Meeting on
Mathematical and Experimental Physics, where parts of this  work have been presented,
 and especially to Leo Garc\'ia-Col\'in and Francisco Uribe for
discussions. I.V.K. acknowledges stimulating discussions with Miroslav Grmela.
 Finally, it is our pleasure to thank
 Misha Gromov and
 Hans Christian \"Ottinger for stimulating discussions.

S. Ansumali, P. Gorban, V. Ilyuschenko, and L. Tatarinova help us to prepare the manuscript.

\end{document}